\documentclass{article}

\usepackage{arxiv}

\usepackage[utf8]{inputenc} %
\usepackage[T1]{fontenc}    %
\usepackage{hyperref}       %
\hypersetup{hidelinks}
\usepackage[hang,flushmargin]{footmisc} %
\usepackage{url}            %
\usepackage{booktabs}       %
\usepackage{amsfonts}       %
\usepackage{nicefrac}       %
\usepackage{microtype}      %
\usepackage{lipsum}
\usepackage{fancyhdr}       %
\usepackage{graphicx}       %

\usepackage{macros/aaai-macros}
\usepackage[numbers,sort&compress]{natbib}
\usepackage{xurl}
\usepackage{hyperref}
\hypersetup{hidelinks}

\usepackage{tabularx}
\usepackage{lscape} 
\usepackage{enumerate}
\usepackage{amsmath}

\usepackage[inline]{enumitem}

\usepackage{pgfplots} 
\pgfplotsset{compat=1.18}

\AtBeginDocument{
  \fancypagestyle{firstpagestyle}{
    \fancyhead{}%
    \fancyfoot[C]{}%
  }
  \fancyhf{}
\pagestyle{fancy}
\fancyhf{}

\newcommand{\runningtitle}{Norms, Rules, and Moderation in AI-Generated Sexual Content Communities}
\newcommand{\runningauthors}{Qin, Mink, Redmiles}

\fancyhead[L]{\sffamily\footnotesize\ifodd\value{page}\runningtitle\fi}
\fancyhead[R]{\sffamily\footnotesize\ifodd\value{page}\else\runningauthors\fi}
\fancyfoot[C]{}

  \fancyfoot[C]{}%
}
\makeatother

\title{``I Thought You Were The Uncensored Place'':\\Norms, Rules, and Moderation in\\AI-Generated Sexual Content Communities}

\author{
Lucy Qin\footnotemark[1] \\
  Georgetown University\\
  \texttt{lucy.qin@georgetown.edu} \\
   \And
   Jaron Mink\footnotemark[1]\\
Arizona State University \\
\texttt{jaron.mink@asu.edu}
     \AND
  Elissa M. Redmiles \\
  Georgetown University\\
  \texttt{elissa.redmiles@georgetown.edu} \\
}

\begin{document}
\maketitle

\begingroup
\renewcommand{\thefootnote}{}
\footnotetext[1]{* \textbf{Both authors contributed equally} to this work, share first authorship, and reserve the right to adjust ordering on personal documents.}

\footnotetext{A version of this work appears in the proceedings of {the 2026 AAAI/ACM Conference on Artificial Intelligence, Ethics, and Society (AIES)}.}
\endgroup

\begin{abstract}
As AI-generated sexual content (AIG-SC) is increasingly produced, online communities have emerged to support creators' needs.
To understand whether and how community governance attempts work to prevent abuse while supporting free expression, we interviewed 24 members and moderators of large AIG-SC online communities (10,000+ members) with stated rules against creating and sharing abusive content (e.g., AI-generated CSAM).
Through in-depth interviews, we \edit{offer insight into}:
(1) \edit{how and why these communities form}; 
(2) implicit community norms; 
(3) explicitly stated rules---and their operationalization via content moderation; 
and (4) tensions between community values and moderation that leave space for abusive behavior. 

Our findings reveal a complex picture: while many creators and communities have personal boundaries against abuse, advice and resources for creating \textit{any} form of AI-generated sexual content are accessible to users regardless of their intentions.
Further complicating community moderation are norms that center anti-censorship and non-judgment, which leave moderators to justify their actions using the limits of the law and terms of service. We end by reflecting on the ways in which technical, community, and legal governance may most effectively mitigate the production of abusive content.

\end{abstract}

\vspace{1.5\baselineskip}

\noindent\fbox{%
  \parbox{0.98\columnwidth}{%
    {\textbf{Content Warning:} This paper discusses sexual violence, non-consensual intimate imagery, CSAM, and quotes from people who made non-consensual intimate imagery via AI.}
    }
  }

\newpage
\section{Introduction}
Like prior forms of technology-mediated sexual content~\cite{fullerPornWarsSerious2019}, questions of whether AI-generated sexual content (AIG-SC) is produced or consumed ethically have become a focused societal concern. \edit{The public availability of retrainable open-weight models (e.g., SDXL, Flux), model repositories (e.g., \civitai), and AI jailbreaks have enabled creators to generate virtually {any form} of sexual content~\cite{thielGenerativeMLCSAM2023, 
emanuelmaibergAIPornMarketplace2023, minkUnlimitedRealmExploration2026, hawkinsDeepfakesDemandRise2025}.}
While some may legally create AIG-SC to explore sexual, creative, and technical interests~\cite{minkUnlimitedRealmExploration2026}, others use these same technologies to harm others in AI-generated non-consensual content (AIG-NCC\footnote{We use the following acronyms to refer to different types of AI-generated content. AIG-NCII refers to AI-generated non-consensually created intimate imagery (sometimes referred to as ``sexualized deepfakes'' or ``deepfake pornography'') that depicts a known individual’s likeness. AIG-CSAM refers to AI-generated child sexual abuse material. AIG-NCC refers to AI-generated non-consensual content, which we use as an umbrella term to include both AIG-NCII and AIG-CSAM.}) via non-consensual intimate imagery (AIG-NCII;~\citealt{timmermanStudyingOnlineDeepfake2023,gibsonAnalyzingAINudification2025,hanCharacterizingMrDeepFakesSexual2025, umbachAIgeneratedImagebasedSexual2026}) and child sexual abuse material (AIG-CSAM;~\citealt{wei2025utterlyillprepared}).

In response, private and public institutions have taken measures to attempt to reduce AI-enabled sexual violence.
Federal institutions and private institutions have appropriately taken down communities dedicated to enabling harm, such as the MrDeepFakes forum~\cite{hanCharacterizingMrDeepFakesSexual2025,timmermanStudyingOnlineDeepfake2023} and Telegram channels~\cite{burgessMillionsPeopleAre2024, tsuchiyaLargeScaleStudyTelegram2026}.
Similarly (and perhaps in light of irresponsible AI developer behavior;~\citealt{katecongerElonMusksGrok2026}), AI developers have largely prohibited the generation of all sexual content, abusive or legal, within their platforms.
While lauded by some, digital civil rights groups 
are increasingly concerned about whether \edit{lawful} (sexual) \edit{expression} is being censored and deplatformed~\cite{EFF_Flawed_Take_It_Down_Act_2025, centerfordemocracytechnologyTAKEITAct2025}.
At the same time, others are weaponizing lawful expression to protect NCC-producing `nudify' apps~\cite{associatedpress2026_elonSuesMN}.

In the wake of these rapid legal shifts and deplatforming of sexual content by mainstream AI-platforms, numerous peer-organized online AIG-SC communities have emerged.
While many communities may clearly harbor deplatformed abusers~\cite{cuevas2026deepfake},
other large communities posture as refuges from censorship and for free sexual expression, yet are against AIG-NCC.

However, it is \edit{unknown} whether, and if so how, these communities prevent and/or enable abuse. 
\edit{This is especially important given that there are  no defenses currently available that guarantee AI-based image-generation technology cannot be used to generate AIG-NCII~\cite{chandraReducingRisksPosed2024}. %
Prior work has found that community-supported norms and rules are effective deterrents against cyber-crimes, and are essential (alongside legal and technical interventions) within a comprehensive response to cyber-crimes~\cite{williams2006virtually}.
While prior work has investigated what, how, and why \textit{individual AIG-SC creators} make content~\cite{minkUnlimitedRealmExploration2026},
to our knowledge, none have analyzed \textit{AIG-SC communities} and whether (and how) they support or discourage AIG-NCC creation.}

To analyze behaviors supported within AIG-SC communities,
we draw on analytical methods from organizational psychology and institutional economics.
\textit{Group formation}~\cite{harris2019joining} finds that the process by which communities are created often determines group behaviors, composition, goals, and outcomes; however, %
it is unknown:
\begin{enumerate} [label=\textbf{RQ\arabic*},labelindent=0pt, leftmargin=*, align=left]
\item
How and why do AIG-SC communities form, and their members join? 

\label{rq:lifecycle}
\end{enumerate}

Once groups are established, Ostrom's  
institutional development and analysis framework~\cite{ostrom2011background}
notes that \textit{group (a) norms, (b) rules, and (c) governance} are essential in predicting and governing their behavior.
However, within AIG-SC communities, it is unknown:
\begin{enumerate}[label=\textbf{RQ\arabic*}, resume,labelindent=0pt, leftmargin=*, align=left]
\item What norms dictate interactions, and what beliefs are professed among members?\label{rq:interactions}
\item What personal commitments/rules do (a) individuals and (b) communities use to govern their actions?\label{rq:boundaries}
\item What tensions, failures, or challenges exist within community-based AIG-SC governance? \label{rq:enforcement}
\end{enumerate}

To answer these questions, we conduct 24 in-depth interviews with members of three large (10k--50k+ members) AIG-SC Discord communities (two image-focused, one text-focused) that publicly prohibit AIG-NCC.

\section{Related Work}

\noindent\textbf{AIG-SC.} With the growing availability of AIG-SC generation services and resources on training customized AIG-SC models, it is now possible to create seemingly endless amounts of personalized sexual content via specialized AIG-SC websites~\cite{lapointePresentFutureAdult2025},
online forums~\cite{doringExperiencesAIGeneratedPornography2025},
or even with assistance from other AIG-SC creators~\cite{minkUnlimitedRealmExploration2026}.
Through interviews with AIG-SC creators~\cite{minkUnlimitedRealmExploration2026} and qualitative analysis of online AIG-SC applications~\cite{lapointePresentFutureAdult2025, lapointeGovernanceAIgeneratedPornography2026}, multiple studies have examined how advancements in AI have 
enabled many to explore their sexual, creative, and technical interests, as well as enabled 
the production of AIG-NCII. 
Discussions and motivations for creating AIG-SC can be found intermingled with clear intentions to make AIG-NCC of celebrities, work colleagues, friends, and ex-partners~\cite{doringExperiencesAIGeneratedPornography2025}.
While some creators within these groups may openly condemn AIG-NCC, others rationalize such abuse~\cite{minkUnlimitedRealmExploration2026}.

Thus far, studies on the governance of AIG-SC spaces have been minimal; to our knowledge, only \citet{lapointeGovernanceAIgeneratedPornography2026}'s  analysis of AIG-SC terms of service, privacy policies, and related documents of AIG-SC (and AIG-NCC) generation and hosting sites has found that although policies against AIG-NCC exist, they largely serve as legal protection for site owners.
While \citet{lapointeGovernanceAIgeneratedPornography2026} also found that moderation appears to rely on user-reporting,
no work has actually investigated how this moderation occurs in practice. 
Furthermore, no work has investigated whether AIG-SC creator communities establish normalized practices around their own content creation, the content creation of other members, or whether and how self-governance occurs.

\header{AIG-NCC} Similar to AIG-SC, AIG-NCII and AIG-CSAM are becoming easier to create through nudification ~\cite{gibsonAnalyzingAINudification2025} and face-swap apps~\cite{daffalla2026dual},
model repositories~\cite{maibergCivitaiBanReal2025, hawkinsDeepfakesDemandRise2025, newtonMyNSFWVideo2020} and AI companions~\cite{brighamExaminingRisksAI2026}. 
AIG-NCC is even permeating mainstream platforms
via listings on Fiverr~\cite{dawoudUndergroundMainstreamMarketplaces2026}
and AIG-NCC-capable chatbots on X~\cite{chanMusksGrokChatbot2026,wilsonHundredsNonconsensualAI2026}.

AIG-NCC creation has also been found to be supported through online communities. Prior to its shutdown, ~\citet{hanCharacterizingMrDeepFakesSexual2025} documented that the MrDeepFakes forum acted as a hub for information to create AIG-NCC, with skilled members teaching others. %
Although MrDeepFakes has since been shut down, AIG-NCC communities persist on platforms such as Reddit and 4chan~\cite{medeirosCharacterizingResourceSharing2026,cui2026celebrities}. 
In response, researchers have investigated aspects pertaining to AIG-NCC, including:
interventions to prevent AIG-NCC generation~\cite{baSurrogatepromptBypassingSafety2024, cretu2025evaluating}, people's preparation for and handling of AIG-NCC~\cite{pendse2025testing, wei2025utterlyillprepared}, 
and offenders' attitudes and behaviors~\cite{flynnSexualizedDeepfakeAbuse2025,umbachPrevalenceImpactsImageBased2025, umbachNonconsensualSyntheticIntimate2024,brighamViolationMyBody2024}.

\header{Moderation of Online Communities, Sexual Content, and AI Content}
While several studies have investigated aspects relevant to AIG-SC communities, none explicitly address the unique legal, community-based, and safety concerns of this work.
Several works have considered the difficulties volunteer moderators face within platforms generally, and have been centered around online forums~\cite{li2022all},
live streaming~\cite{wohn2019volunteer} and the emotional toll and burnout that often follows~\cite{schopke2024volunteer}.
Other work has investigated how the moderation of content deemed as sexual disproportionately impacts vulnerable populations such as sex-workers~\cite{tiidenberg2021sex, blunt2021automating}, queer communities~\cite{zolides2021gender,taylor2025straightening, mayworm2024misgendered}, and vulnerable populations broadly~\cite{haimson2021disproportionate}. 
Other work has more specifically focused on moderating \emph{human-originating content} from \emph{AI-generated content}~\cite{10.1145/3772318.3790415, mink2022Deepphish, lloyd2025there}, finding that such moderation is challenging and can lead to incorrect, biased decisions against real users~\cite{mink2024Moderation}. 
However, no work investigates how moderators of AI-focused communities attempt to distinguish between \emph{AIG-NCC} and \emph{AIG-SC}, both of which are AI-generated and of a sexual nature.

\section{Methodology}

We interviewed 24 AIG-SC creators between April to August 2025 on 
(1) their background with sexual content and AI, 
(2) their creation process, 
(3) their participation within any communities centered on AIG-SC, 
(4) any norms/boundaries they had pertaining to content creation, 
and (5) their content or resource sharing practices. 
Procedures for ensuring safety for all researchers, participants, and potential victim-survivors are detailed in Section~\ref{sec:ethics}.

\header{Recruitment and Participants} 
We created a list of active Discord communities using relevant search terms that focus on generating AIG-SC and explicitly prohibit AIG-NCC.\footnote{Despite this, we prepared for, and encountered, AIG-NCC offenders (see Section~\ref{sec:ethics}).}
We then reached out to the moderators to confirm their approval to interact with and recruit from their community.
With approval, we distributed our recruitment information and responded to any requests for clarification from the community. 
Three Discord communities and their associated moderators participated. \communityA~($10$k--$20$k members) and \communityB~($50$k+ members) focused on visual content  (e.g., images, videos); \communityC~($10$k--$20$k members) supports users with text content (erotic fiction, role-playing). Demographic details are documented in Appendix~\ref{appendix:demographics}.
From each community, we interviewed between 2 to 14 members. Participants were offered $\$40$ USD.

Interested participants were provided a screener survey
in which they answered questions on (1) their use of AI for generating sexual content, (2) their demographic background. 
From this pool of submissions, participants were invited to schedule a 90-minute interview with the research team over an end-to-end encrypted call. When reviewing responses to select participants, we prioritized diverse use cases for AIG-SC creation.
Overall, we collected 30 hours and 36 minutes of interview data; interviews lasted 76 minutes on average.
In most cases, participants were paired with an interviewer with a matching gender presentation~\cite{harling2019influence,wilson2002effects}.

We associate participant identifiers with the  server they were recruited from ``[A/B/C]XX''; to also associate abusive behavior
and beliefs, we attach \osymbol{} to the participant identifier of those who disclosed generating AIG-NCC (e.g., \Peleven).

\header{Analysis}
We conducted reflexive thematic analysis~\cite{braun2006using} and followed an inductive coding process. 
This began with each author reviewing a selection of interview transcriptions to explore themes within the data. 
Through this process, each author created an independent preliminary codebook. 
They then met, discussed themes that emerged, and merged their codebooks into a consolidated version that was used to code an initial set of three interview transcriptions that were triple-coded by all three authors.
Afterwards, the authors met to discuss their coded transcriptions and resolved any discrepancies that emerged. This also resulted in minor revisions to the codebook. The authors then coded two more interview transcriptions using the codebook. They then met to further resolve discrepancies and refine the codebook. This discussion produced a finalized codebook (see~\cite{osf-aigsc-norms}). 
The three authors then divided the remaining transcripts and independently coded them.
While we report on commonalities in what we observed,  we
emphasize that each participant’s values and boundaries are deeply personal
and vary widely. We encountered a range of perspectives on AIG-SC and advise against treating AIG-SC creators as a monolith.

\begin{comment}

We conducted reflexive thematic analysis~\cite{braun2006using} and followed an inductive coding process. 
This began with each author reviewing a selection of interview transcriptions to explore themes within the data. 
They then met, discussed themes that emerged, merged their codebooks, and applied that codebook to three interview transcriptions that were triple-coded by all three authors. After discrepancies were resolved, this was then repeated with two more transcripts.
The three authors then divided the remaining transcripts and independently coded them.
We report on the thematic commonalities in what we observed. %
\end{comment}

\header{Limitations}
We do not claim generalizability of our participants or of all AIG-SC communities.
We only selected communities with rules against AIG-NCC. Those who knowingly offend may be less likely to participate.

\header{Positionality}
We conduct this work to enable prevention of sexual violence, while preserving individuals' right to sexual expression. 
The team believes in a sex-positive approach to addressing and governing the production of sexual content. %
Our team contains members who hold nearly a decade of experience researching security, privacy, and safety concerns related to image-based sexual abuse, sexual content creators, and AIG-NCC.
The team includes researchers who have experienced sexual abuse. {None of the team actively participates in an AIG-SC community.}
We strongly believe in firm policies and protections against all forms of sexual violence, while also supporting individuals' right to freedom of expression, including sexual autonomy.

\subsection{Ethical Considerations}
\label{sec:ethics}
Throughout this research, the research team engaged in ongoing discussions on how to ensure this work could be done ethically and responsibly.
This was considered both through our research methods with participating stakeholders (i.e., the study participants and interviewing researchers), and the expected impacts this work may have, including how it may impact non-abusive sexual content creators, those impacted by labor displacement and intellectual or labor theft (e.g., from sex workers and content creators), and victims-survivors of AIG-NCC.
We now discuss the procedures we followed to minimize risks.
All procedures were approved by the IRB of each author's institution.

\subsection*{Research Procedures}
While the primary focus of our research concerned how members and moderators interacted and governed themselves in AIG-SC communities with explicit rules against the sharing of AIG-NCC (i.e., AIG-NCII and CSAM), we prepared for the possibility that members we interviewed may be offenders of sexual violence. 
To account for this, we created a set of procedures for ensuring the entitled (but limited) protection of all research subjects,
as well as procedures we would follow if a participant revealed that they had perpetrated, or intended to perpetrate, sexual violence.

\header{General Procedures For Participant Protection}
For the general participant procedure, we follow best practices to ensure participant protection.
All participants were recruited from large communities of over 10,000 members.
Our research team received permission from moderators explicitly before interacting with any members or posting content within their community.
Every participant was provided instructions for how to email and conduct the interview anonymously.
Participants could choose to be interviewed over video, audio, or text.
Prior to recording, each participant was reminded of the study's purpose and asked to reaffirm consent before being recorded.
Participants could end the interview at any point, skip any question, or retract their data after the interview, though none did.

\header{Procedures for Disclosures of Sexual Violence} 
While we intentionally limited our recruitment to members of communities that explicitly prohibit the sharing of AIG-NCC, we prepared procedures in case a participant disclosed AIG-NCC creation during an interview.
To build our procedures, we referenced two key sources.
\textit{First}, established PIs and researchers who have experience either
working in a sub-domain in which participant retaliation was a consideration, 
or working in a field such as psychology and/or criminology in which established norms for conducting qualitative research with offenders of sexual violence existed.
\textit{Second}, we extensively utilized the recommendations from established guidelines for interviewing offenders of sexual violence.
Specifically, we considered how best to protect researchers from harm during these interactions and minimize harm to current or future victim-survivors. %

Despite our recruitment methodology, three participants disclosed that they had created AIG-NCII during the course of their interview.
We now discuss the procedures we have prepared for handling such disclosures.

\smallskip
\noindent
\subheader{Limited Confidentiality}
To prioritize the current and future safety of victim-survivors, we follow established guidelines~\cite{hearn2007background, sexual2012ethical, sikweyiya2011perceptions, cowburn2005confidentiality}
and implement limited confidentiality within our study procedures and consent process.
If a participant discloses a clearly identifiable criminal offense in which they (1) create CSAM, or (2) publicly upload AIG-NCC of an identifiable victim, or state an intention of imminently doing so, we would end the interview (if active), report the activity to NCMEC cybertip line~\cite{missingkidsHome} (for CSAM) and a platform authority (if publicly uploaded).
This limited confidentiality was established within our approved IRB procedures and similarly disclosed when participants' consent was obtained.
Following best practices as researchers~\cite{cowburn2005confidentiality}, we do not act as law enforcement and do not attempt to elicit more details than necessary to complete our research, and we do not attempt to elicit information that would require breaching confidentiality. 
Ultimately, we were required to breach confidentiality for one participant who sent the interviewer their public account over chat during the interview. After the interview concluded, publicly posted NCII of an identifiable adult was discovered, and that content was reported to the platform and removed. %
After our study concluded, moderators from the participating communities were informed of these results, including the existence of offenders of sexual violence who may have been recruited from their community.

\smallskip
\noindent
\subheader{Preventing Condonement of Sexual Violence}
To prevent normalization of AIG-NCC behaviors, we follow established guidelines~\cite{sexual2012ethical, hearn2007background}, and ensure that interviewers do not condone acts of sexual violence either through verbal or non-verbal cues to any participant.
To protect researchers and prevent confrontation, interviewers did not provide any opinions on offending behavior, but did question the appropriateness of such behavior if minimized by the participant~\cite{sexual2012ethical}.
As a result of our interview, participants appeared to be aware that the creation and distribution of the AIG-NCC is unacceptable.
\Pfortysix{}, who created AIG-NCII of a public figure, later denied ever using AI to make sexual content; another moderator who did not engage in AIG-NCC added significantly more restrictions to their communities' rules, including those relating to AIG-NCC.

\smallskip
\noindent
\subheader{Protection of Researchers}
All researchers on this study were aware that the work carries a non-negligible risk and openly discussed this; in addition to being a potential target of AIG-NCC, potential community backlash is unfortunately a real possibility~\cite{doerfler2021m}.
All researchers were provided access to a mental health consultation throughout the project.
As this was the team's first time interacting with these communities, and out of an abundance of caution, we attempted to minimize power dynamics by not including any persons in degree-granting programs (e.g., PhD, Masters, Undergraduate).
Our team fully believes that including students is necessary for academic mentorship and establishment of norms but made this exclusion out of extreme caution in interacting with an unknown community.
We have since established trust with moderators in AIG-SC online communities and are conducting follow-up work that includes students.

\section{How and Why Do AIG-SC Communities Form, and Their Members Join? (\ref{rq:lifecycle})}
In asking AIG-SC moderators and members about how they became involved, we surface how/why their communities are created and why/how members join.

\header{Discovering Communities}
Search engines and general-purpose LLMs (e.g., ChatGPT) helped several participants discover AIG-SC online communities. 
For instance, \Pthirtynine{} was curious about AIG-SC and naturally thought to \pquote{Google like anything just to see like what is being done in the space.}
Additionally, participants such as \Pfive{} noted that posted tutorials and guides often linked to AIG-SC communities:
\pquote{I found a GitHub page that explained how to use [a video generation model] locally... I see [their community] was linked there in that repository}
For those focused on text-based roleplaying or erotic fiction generation, safety filters on general purpose LLMs (e.g., ChatGPT, Claude), led them to seek out AIG-SC communities. 
For example, \Pseventeen{} expressed that they were \pquote{frustrated by forced restrictions that do not actually stop AI from being used in harmful ways.}

\header{Searching for and Exchanging Technical Knowledge} 
To support technical advice-sharing, two participants created AIG-SC communities on Reddit and Discord to centralize relevant information.
For example, \Pthirtytwo{} had searched for reliable information in multiple sub-Reddits but found that they were mismanaged or contained spam. Afterwards, \Pthirtytwo{} \pquote{decided to do it myself...
and made sure to build [the community] in a way that everyone can have access to every tool they need to enjoy an uncensored experience with AI.} 
Similarly, \Pthree{} shared the goal of cultivating spaces for open collaboration in order to improve tools and techniques for AIG-SC: \pquote{my focus is on...
getting people interested in the building, the research aspect of it, collaborating on best practices, exploring what the capabilities are}~(\Pthree).

Members within these AIG-SC communities found that the quality and availability of technical advice on AIG content was higher in AIG-SC spaces \edit{than in other online communities that discuss generative AI}, and even drew in those not initially interested in sexual content: \pquote{I had joined many different Discord servers for people that are talented in this area. And it just so happened that the most talented people I’ve been able to find were in the Not Safe for Work space}~(\Ptwelve{}).
In order to find the best advice to support their AIG-SC experimentation, \Pfortyseven{} also \pquote{joined a lot of Discord groups.}
These AIG-SC communities were described as pushing technical boundaries. \Pnine{} notes that the community they participate in often tackles generation tasks not yet approached elsewhere, \pquote{mixing concepts together, coming with more complex images, that is the thing that you see.}

What started as technical help quickly turned into social bonding for several members:
\pquote{I would have them walk me through it... and then I’d forget everything and have them walk me through it again the next day... 
and it ended up being like over a month of consistently talking like every single night with these guys}~(\Ptwelve).
After receiving this assistance, participants reported feeling a responsibility to give back and assist others: \pquote{I put effort into helping them with what I can do, because I like this kind of sharing the knowledge} (\Pten).
Some members, such as \Pnine, described these AIG-SC spaces as 
\pquote{a two-way helping hand,} noting that from these experiences, their peers consisted of \pquote{good people. [chuckles] They help everyone.}

Beyond social connections, others felt that AIG-SC communities may also be useful in forging professional connections.
For instance, \Pthirtyfour{} notes that some start-up ideas have emerged from Discord-based collaborations:
\pquote{The best thing that came out of it was the cool technical collaborations... 
maybe I’ll find something where I could get a job somehow through some of this}~(\Pthirtyfour).

\header{Rejection and Deplatforming}
\label{sec:joining-communities:rejection}
Participants created or joined AIG-SC communities on Discord because they were rejected elsewhere. %
They described varying tolerance towards AIG-SC in other peer-organized online communities. 
For example, \Pten{} creates sexual content that depicts an anime character and \pquote{the community around the character... they don't like [AI]}. 
When they previously shared their AI-generated outputs, \pquote{people start to complain.}
Participants most often experienced rejection in communities with a tradition of artistic patronage: 
\Pthirtyseven{}, an artist within the furry community\footnote{A fandom in which members depict alternative identities, often through commissioned art~\cite{austin2023identity}.}, mentioned hiding their use of AI as
\pquote{artists they tend to have a very negative viewpoint on AI, and if I were to share that information that `hey, I enjoy creating AI images,' a lot of them would be like, `okay, I don't want to talk to you'...I can't risk severing those connections.}

Conversely, more general online spaces (e.g., AI art communities) that participants frequented often prohibited sexual content. As such, AIG-SC communities served as a \pquote{very niche special interest group} (\Pthirtyfour) for the convergence of these interests.
Here, participants found consenting audiences with shared values:
 \pquote{porn is usually something you keep to yourself but here you’re sharing it} (\Ptwelve).
Even in offline environments, these communities served as one of the few, or only, places in which they felt comfortable discussing AIG-SC:
\pquote{I used to have a therapist... I was really open to her about everything else...
but I never told her about [this]...
I wasn't sure if she would understand}~(\Pfortyseven).

However, even though these communities emphasize open-mindedness (see \textbf{Norm \#3}) they still can be discriminatory:  \Pthirtyfour{} mentioned
that queer AIG-SC content was often marginalized,
\pquote{they’re into some extreme kink...
but then they see two dicks, and they’re like, `Oh my God!'...
it's kind of funny how homophobia can present itself even in [the] extreme kink underground.}
For this reason, \Pthirtyfour{} and other queer members decided,
\pquote{let’s go to [a new AIG-SC community]... an all-M4M community.}

Others found new AIG-SC communities when their prior ones were banned or moderated.
\pquote{[My prior community] started an anti-not-safe-for-work crusade...
Everybody scattered to other [AIG-SC] sites}~(\Pseventeen{}).
Similarly, as the creator of Community A, \Pthree{} mentioned that they wrote an AIG-SC article and \pquote{it got deleted from Reddit, which is what led to me creating the Discord.}
Others noted they
\pquote{didn’t like the way that [a prior community] was being moderated... 
All the old stuff that I posted got deleted... I was like, `Fuck that group!'}~(\Pthirtyfour{}).
Occasionally participants described deplatforming events triggered by policy changes due to public scrutiny: 
\begin{quote}
\pquote{Nobody knows who, but someone went to TikTok, and showed... everything...
people massive message[d] [the AI platform]...
and [they] immediately [removed AIG-SC]... everybody was so pissed... 
a lot of them just went to [a different platform]}~(\Pfortythree{}).
\end{quote}

\section{Community Norms (\ref{rq:interactions})}
\label{sec:interactions-and-norms}

\edit{Similar to existing works that identify social norms of online communities via qualitative interviews~\cite{10.1145/3415226, rashidi2020s, siitonen2007social}, we identify norms by analyzing %
behaviors participants report as being allowed, disallowed, and intentionally sanctioned by their communities, which indicate the existence of, and often define, upheld community norms~\cite{liefbroer2010bringing,opp2001norms}.
Through this, we find that AIG-SC communities often hold beliefs that (1) stand against perceived censorship, (2) support low-cost communal learning, and (3) reinforce judgment-free spaces, which are essential in advancing the expertise and infrastructure needed for AIG-SC creation.}

\label{subsec:upskilling}

\header{\textit{Norm \#1: ``AI Needs to be Uncensored''}}
\label{subsec:norm1}
Participants wanted to create, and enable others to create, AIG-SC limited only by the bounds of their own imagination: \pquote{the whole point of the Discord, of the community, is trying to make the generative AI models as diverse as possible, as customizable. Like, there’s no limitations} (\Peleven). Corporate and regulatory governance over AI content generation were a large concern, which they viewed as censorship:~\pquote{[Companies] censor.
They have the power over what you can do, and what you cannot do}~(\Pten{}).
Such feelings were inflamed when credit card payment processors rescinded from \civitai, a popular model-hosting platform,
due to the prevalence of models specialized to create NCII~\cite{maiberg25_civitai_cutoff}. Such action was viewed by several participants as AI censorship.

Those with text-based use cases often sought out AIG-SC communities after encountering safety filters on general LLMs (e.g., on ChatGPT or Claude):
\pquote{I got very annoyed at the sanitization of ChatGPT from filters, so I searched for NSFW AI communities}~(\Pthirtysix). 
Such participants were particularly sensitive toward censorship concerns. For instance, \Pseventeen{} expressed that they were \pquote{frustrated by forced restrictions that do not actually stop AI from being used in harmful ways.}  
Therefore, such participants intentionally sought out others who were interested in circumventing filters in order to enable sexual content creation: \pquote{if you’re going to look for ways to get past [safety filters], you’re going to find out that other people have had these questions also and bit by bit, you end up in the sorts of places where people are wondering this in general} (\Psixteen).

As a result of these censorship concerns, participants act to make AIG-SC communities havens for creative freedom with AI.
People like \Peleven{}, who became interested in the AIG-SC communities after finding that it \pquote{wasn’t some group that hell bent on making AI porn. It was just like, these guys are making [AI] uncensored} believes that \pquote{the whole purpose [of AI is that] you have the freedom to make what you want.} 
Likewise, \Pseventeen{} describes that the AIG-SC community includes 
\pquote{people who, like me, are frustrated by forced restrictions that do not actually stop AI from being used in harmful ways} and their belief that \pquote{if AI is really going to reach its full potential, where people who lack creativity but have great ideas are able to actually bring those ideas to reality... AI needs to be uncensored.} 

\edit{Limiting sexual content generation}  was perceived as the beginning of a slippery censorship slope:
\pquote{if you can block people from [making sexual content] then you can pretty consistently get [AI] to stop doing other stuff}~(\Pfortytwo{}). \Pfortytwo{} draws parallels with prior forms of controversial art and creative expression that have been previously suppressed by institutions: \pquote{Film history is full of like examples of people who made subversive things that were like horrible, or that were called various things but are still meaningful pieces of art.} This discourse echoes academic analysis of AI-based censorship of artistic nudity~\cite{riccio2024exposed}, but also echoes arguments from AIG-NCC creators that the abusive content they generate is artistic~\cite{hanCharacterizingMrDeepFakesSexual2025, flynnSexualizedDeepfakeAbuse2025}.  

\header{\textit{Norm \#2: ``Knowledge should ALWAYS be free''}} 
\label{subsec:norm2}
Participants characterized AIG-SC community culture as driven by collaborative knowledge exchange of technical resources to advance the state-of-the-art.
Although some members use AIG-SC communities to create products for monetization, free \textit{information exchange} is a critical cultural value, as reflected by the amount of free, collectively-authored technical resources offered in AIG-SC communities and the availability of free peer assistance. Monetization was a divisive issue: as \Pthirtyseven{} described, 
\pquote{Some people prefer it kind of not monetized, and others are more than happy to monetize the content and sell it for a profit. 
That's often the divisiveness that I see...Most of the time, it's over money.} \Pthirtytwo{} noted that in another community
\pquote{people SOLD a jailbreak...that [behavior] goes against everything I believe and stand for...knowledge should ALWAYS be free.}

Since participants valued free information exchange, concerns around monetization often stemmed from the belief that others were being scammed by private companies offering low-effort AIG-SC generators that relied upon the same free resources available to creators. %
This sentiment led 
 \Psixteen{} to provide the following characterization of those profiting from AIG-SC content generation products: 
\pquote{They're liars. They’re grifters trying to make money off of hooking up cheap models with simple instructions...
 And then getting people to pay...
It is a terrible deal.}

A shared cultural value around free information exchange also results in norms around free data collection and sharing.
No participant mentioned paying for data (to e.g., fine-tune AI models) or obtaining consent to use it.
As a whole, it is not clear where participants collected their data from, though in a few cases we observed that they collected content from social media and adult sites, such as \pquote{Reddit and XHamster} (\Pthree). This is, unfortunately, in line with findings that social media and adult sites are being scraped nonconsensually to train AIG-SC models~\cite{emanuelmaibergBoomingAIPimping2024,princessacintaqiaStopNonconsensualUse2025}. 
\edit{Participants rarely engaged deeply} with the ethics of nonconsensual data use. Rather, a few lamented that outsiders joined their community periodically to advocate against the theft of training data or business from artists. For example, \Pnine{} explains that sometimes \pquote{there are some guys that just enter [the community] and say, `stop using this, this, this. Stealing artists' works.'}

\header{\textit{Norm \#3: Don't Judge, ``Don't Ask, Don't Tell''}}
\label{subsec:socialization}
\label{subsec:norm3}
Given the stigma around sexual content and the ostracization participants may have received (see Section~\ref{sec:joining-communities:rejection}), they expressed a strong desire for nonjudgmental spaces that allow them to discuss and share the content they produced freely: 
\pquote{You're doing way more harm to this world by stifling sexual expression... sexual stimulation is a form of self-care}~(\Pfortyone). 
\edit{As noted by several moderators, their communities were built upon norms of non-judgment that were often essential for encouraging the community participation required to develop AIG-SC creation expertise:
\pquote{we try to set up a place for people to learn how to use the tech so they can explore their niches and for people to share their art without fear of judgment} (\Pthirtyone{}).} 
This nonjudgmental spirit enables members to find interested audiences and explore sexual \textit{niches} (e.g., content relating to a specific kink). For example, \Peleven{} noted that, \pquote{there's all these people contributing to the knowledge of how to make photorealistic cat girls.}\footnote{Cat girls are a common archetype in anime, typically depicting a human woman with cat-like features (e.g., cat ears).}.
While members may not all share the same sexual interests,
\pquote{there’s different people into different stuff. But it is generally pretty supportive} (\Pthirtyfour).
As \Ptwelve{} described, \pquote{it's people genuinely sharing what they're into and that might be shocking to some} but \pquote{people are really open-minded. And that’s actually really nice to just have people you can share whatever with, especially stuff that’s so intimate like this.} Accordingly, participants tended to deliberately ignore others' implied actions, as long as the resulting content was not shared: \Pfortytwo{} notes that \pquote{when it comes to like stuff that we ourselves generate it's usually kind of said `don't ask, don't tell'}; however, if content is shared that is \pquote{very obviously undefendable, we will just complain about it and then [a moderator] bans them immediately.} When discussing content with darker, more violent themes, \Ptwentysix{} commented that, \pquote{a lot of people make some make some pretty gruesome stuff. [laughs] And they don’t share. I’m pretty sure there’s been a lot of skeletons in other people’s closets.}

Several participants, such as \Pnineteen{}, believed that as long as the content is private, it does not matter what is generated:~\pquote{The community in general thinks that whatever you want the AI to write is okay, just keep it to yourself.}
Deliberate ignorance and belief that private generation does not harm others contribute to rationalizations of harm as exemplified by \Pseventeen{} who believed that because content can be generated privately, it is \pquote{not harmful to anyone else.} 
This was a recurring sentiment expressed by others, \pquote{AI in of itself is not harmful or hurtful. It's the people that use it}~(\Pthirtyseven). %

In combination with other norms that center creative freedom against censorship and free information exchange, the \textit{``don't ask, don't tell''} mentality can lead to notions of futility around combating harm. 
\Pten{} vocalized that their country has had
\pquote{a lot of incidents lately of students making naked images of their other female classmates... Or people making underage content...
I don’t believe we have actual control over that. It’s like the side effect of creating this technology, and we will need to learn how to live with that.}

\section{Commitments and Rules (\ref{rq:boundaries})}

\label{sec:norms:moderation}
The availability of community-based upskilling, combined with open models and accessible fine-tuning techniques, enables highly customized content to be created with few safeguards, \pquote{You cannot control what people create. There are free models, and people can create whatever they want with that. There’s no way to stop that} (\Pfive). 
To guide what content they create and share, creators negotiate the community norm of uncensored content generation against their own personal commitments~\cite{sen1977rational}, and explicit community rules~\cite{ostrom2011background}. 

\subsection{Personal Commitments}
\label{sec:norms:commitments}

While some participants describe their personal commitments as \pquote{way less restrictive} (\Pnine) than what is allowed in their communities, others personally opted not to share 
\pquote{anything that I feel is morally wrong... none of it’s illegal [but] 
if there’s anything I feel like, `Okay maybe it’s a step too far in the content here,' I don’t share it.}~(\Ptwentysix{}).

\edit{When asked about their boundaries}, most participants explicitly brought up a personal commitment not to generate underage content (either text or images). 
Some, but fewer, mentioned commitments against creating NCII. To avoid creating NCII, \Pnine{} focused on creating anime AIG-SC models without using images depicting humans, stating that, \pquote{I will not train on someone without them knowing that I'm doing this.... I don’t want to get sued.} 
However, a few individuals excluded content depicting celebrities from NCII, viewing these individuals as more implicitly consenting.

In a few cases participants who take commissions received requests for NCII and expressed a clear boundary: 
\pquote{I’ve gotten requests for commissions for deepfake content, which I’ve absolutely refused... because of my own ethical and moral compass. That’s really where I draw the line}~(\Ptwentynine).
Avoiding NCII creation in practice means potentially incorporating more vetting. As further described by \Ptwentynine{}, they have also been approached by people who want to generate content depicting themselves:~\pquote{[An OnlyFans Creator] wanted me to create, AI content of [herself]... she sent me a bunch of reference images}. This places responsibility on creators to verify that the images actually depict the individual making the request. 
Further research is needed on if and how they perform such verification.

While most participants shared a personal commitment not to generate underage content, AIG-SC communities, in general, were described as containing individuals with varying personal commitments. Personal commitments do not necessarily translate into governance of others' behavior. As \Pten{} explains:~\pquote{I don’t do what I don’t like... 
[such as] underage characters. I myself believe it’s disgusting, to be honest. [But] for me, it’s not like if some people I know like it, I will confront them telling it’s bad... that’s their problem, not mine.} Explicit rules, along with moderation, are necessary to establish collective baselines: \pquote{communities often have some kind of ethical guidelines they establish or topics that they request you stay on when posting. Super common}~(\Ptwentyseven). 
In comparison to spaces that conventionally lack moderation, such as 4chan or Telegram, \pquote{Discord is much kinder as a rule; self expression will be more regulated to some degree. Either by the self because it's not anonymous or by the community because they don't tolerate certain types of behavior}~(\Ptwentyseven).
However, the level of depth in community rules and  stringency of enforcement vary.

\header{Comparing Text-Based Communities to Visual-Based Communities}
\label{sec:text-to-vis-communities}
We also found a few qualitative differences in communities' beliefs depending on whether the creator was from a text-based community or a visual-based community.
In particular, text-focused communities 
were much less likely to share content.
As text-content within AIG-SC communities often consisted of highly personalized role-playing scenarios~\cite{juliakiesermanCaughtMafiaRomance2026, minkUnlimitedRealmExploration2026, zengHiddenDesiresExploring2026},
participants reported that role-playing interactions were often
\pquote{a reflection of who you are, what you like, what you don’t like, what you want to explore...
this specific collection and configuration is only going to have full complete relevance to you}~(\Pseventeen).
In some cases, it could even be
\pquote{considered a little bit rude to just like share your responses}~(\Pfortytwo). 
Instead, participants reported that it's much more common to share either jailbreaks as they are often required for AIG-SC text workflows, 
or ways to customize roleplaying scenarios, such as character cards (see Section~\ref{subsec:upskilling}).

Additionally, unlike visual-focused communities, text-focused communities held reservations about the effect of AI-generated erotic fiction on human writers:
\pquote{fic sharing sites like Ao3\footnote{Archive of Our Own (Ao3) is a platform for sharing user-generated stories~\cite{organizationfortransformativeworksArchiveOurOwn}.} are getting flooded with generative AI fanfiction...
flooding what is meant to be a human-creativity sharing space with low-effort content}
(\Pthirtysix).
Perhaps because members of text-focused communities were already writers or involved in communities in which writers depended on patronage, they were often aware of AIG-SC's potential to displace human-created content, and intentionally refrained from sharing AIG-SC. For example, \Pnineteen{} explained that they
\pquote{try to keep the story for myself} to avoid competing with writers.

\subsection{Community Rules}
\label{subsect:community-rules}

Within all three communities, there are explicit, public community rules \edit{(e.g., posted in a designated channel)} that are summarized in Table~\ref{tab:rules_onecol}.
We provide this analysis to give a snapshot of the types of rules that exist within AIG-SC communities and to supplement participant narratives. 
This does not serve as an evaluation of these communities, or AIG-SC communities in general, especially since rules are often in flux and adapt over time:
\pquote{I just put up new rules just earlier today...
we’re going to see the server kind of evolve}~(\Pthree{}).

\newcommand{\alt}{\rowcolor{black!10}}
\begin{table}[t]
\footnotesize
\centering
\begin{tabularx}{.75\columnwidth}{l X X X}
\toprule
Community Rules & A & B & C \\

\midrule
\textbf{AIG-SC} &  &  &  \\
\alt
\hspace{3mm}AI content sharing only & \parti &  &  \\
\hspace{3mm}No use of real likenesses & \full & \full &  \\
\alt
\hspace{3mm}No content that depicts minors & \parti & \full & \parti \\
\hspace{3mm}No obscenity & \full & \full &  \\
\alt
\hspace{3mm}Members must follow Discord's TOS & \full & \full &  \\
\hspace{3mm}Members must follow the & \parti & \full &  \\
\hspace{3mm}TOS/License of AI models & & & \\

\midrule
\textbf{Assets} &  &  &  \\
\alt
\hspace{3mm}No minor enabling tools &  &  & \full \\
\hspace{3mm}No real people & \parti &  &  \\

\midrule
\textbf{Peer} &  &  &  \\
\alt
\hspace{3mm}Duty to report violations &  & \full & \full \\
\hspace{3mm}No kink shaming &  & \full & \full \\
\alt
\hspace{3mm}No discrimination &  & \full & \full \\
\hspace{3mm}No harassment & \full & \full & \full \\
\alt

\hspace{3mm}No political/moral debates &  & \full & \full \\

\midrule
\textbf{User} &  &  &  \\
\alt
\hspace{3mm}Server not liable for posts & \full &  &  \\

\hspace{3mm}Age +18 Self-Attestation & \full & \full &  \\

\bottomrule
\end{tabularx}

\caption{Community Rules --- \textmd{\small
We present communities' posted rules.
\full{} notes that such a rule is present and comprehensive, whereas \parti{} notes that the rule is partially covered, or covered to a lesser extent (e.g., while Community A, B, and C all have rules against content depicting minors, only B also has rules against subjects appears in contexts, like classrooms, that could imply the subject is a minor).
}}
\label{tab:rules_onecol}
\end{table}

\header{AI-generated Output Sharing}
The vast majority of rules across communities focused on AI-generated output sharing.
Text content is less commonly shared for a variety of reasons (see Section~\ref{sec:text-to-vis-communities}). Therefore, rules often pertained to {{visual content}}.
Notably, rules are lacking for asset creation and sharing (see discussion in Section~\ref{sec:rules:assets}).%

\subheader{Age}
Without stringent moderation, a few participants noted that underage content is, in general, \pquote{the most popular theme to develop} (\Pten) across different online spaces (including those not on Discord).
Therefore, some participants emphasized the importance of explicit rules and active moderation of underage content: %

\begin{quote}
   \pquote{[AI] opens the possibility for anyone to create basically anything. 
   [People have] weird fantasies and stuff...that’s [mostly] fine...
    But I have seen a lot of...the child fantasies or pedophilia... 
    it was really, really concerning and really weird seeing images, like, of pictures of girls, small girls... It’s not AI’s fault...
    if [people] have a tool to do something, they are going to do it. So, yeah, I think that it’s better to have a little bit of control on what, at least publicly, [is] available for people to see.}
(\Pfive)
    \end{quote}

In line with participants' personal commitments, all three communities had public rules against sharing content depicting minors. 
Participants noted that the rule against underage content superseded \textbf{\textit{Norms 1 and 3}}. For example, %
\Ptwelve{} explains, %
\pquote{I might not want to see these fat furries with big dicks, or something like that...
But [the community is] going to be like, `Okay, that’s your own right to generate that.' [But the] child stuff... people can agree on, `Okay, we’re not going to generate this.'}
Similarly, \Psixteen{} states that %
\pquote{obviously, anything sexually charged can be a bit sensitive} and they try to be accepting and \pquote{to understand that every person is different...}, but \pquote{there are some things that definitely do cross lines and we’ll tell people [in my community], `No, this is not happening, this is not the place for you.} However, given the inherent ambiguity of AI-generated content, there can be challenges in distinguishing what ``crosses a line'' as we discuss further in Section~\ref{subsec:governance-gaps}.

\subheader{NCII}
Similar to participants' personal commitments, NCII was the next most common content rule mentioned: \pquote{just don't make anything featuring real human beings or children and I think you're fine legally}~(\Ptwentyseven). 
To actualize this, both \communityA{} and \communityB{} had rules against sharing content depicting real people.

\subheader{Obscenity}
In both communities that focused on visual content, there were rules against obscenity such as \pquote{rules against gore and violence} (\Peleven{}). \communityA{} did not include a definition of what qualifies as obscenity but referenced U.S. obscenity laws~\cite{u.s.supremecourtMarvinMillerState1973}, while \communityB{} had rules against the depiction of bestiality, sexual violence, and excessive violence.

\header{Assets}
\label{sec:rules:assets}
In contrast to the development of stringent rules on sharing AI-generated outputs, there is a noticeable gap in rules around asset sharing.

\subheader{Tools Enabling CSAM} 
Generally, there were no stated rules about sharing tools and assets that enable the creation of underage content. In combination with \textbf{\textit{Norm \#2}}'s emphasis on free information exchange (see Section~\ref{subsec:norm2}), the absence of such rules may inadvertently enable the creation of AIG-CSAM and AIG-NCII.
\communityC{} was the only one that had a rule against sharing assets that could enable underage content. Since they focused on text-based use cases, their rules included sharing jailbreaks or prompts that could be used to generate underage content. Community members enforce this by refusing assistance with related prompts and by directly confronting those who violate the rules, \pquote{There are a whole bunch of people now and then who occasionally enter... you will question those people and figure out that, `Huh, these things you’re trying to have it write, these people are wildly underage. We do not approve of this, get out of here'} (\Psixteen). However, this mostly applies to overt requests since members cannot dictate how others use jailbreaks once they are shared. 
To avoid jailbreaks that are overly permissive {as to allow underage content}, 
\communityC{} created their own jailbreaks. Unlike sites like 4chan that \Pfortytwo{} characterized as \pquote{completely unfiltered... and there's nothing that you can do about that}, this community bans those who violate their rules \pquote{as fast as possible}.

\subheader{Data Sharing}
\label{sec:norms:data-collection}
\communityA{} was the only community that had a stated rule against directly sharing visual data that could be used for training~\edit{(e.g., posting data in the Discord server)}. However, \edit{as the server also allows links to datasets and even}
self-hosts 
\pquote{huge files, tons of videos, tons of images, multi-modal sort of stuff...
there’s options where nobody’s going to tell us to take it down} (\Pthree) in order to evade legal and platform regulation, \edit{it casts doubt on whether such rules are genuinely enforced.}

\header{Peer Conduct \& User Requirements}
Several community rules address appropriate conduct.
In particular, several rules instantiate \textbf{\textit{Norm \#3}} %
and explicitly ban
kink shaming (\communityB{}, \communityC{}),
discrimination by race, sex, sexual preference, and other aspects of sexuality (\communityB{}, \communityC{}),
and
general forms of harassment (\communityA{}, \communityB{}, \communityC{}).
In terms of specific behaviors,
only \communityC{} explicitly requires members to ask for consent before sending a direct message, perhaps to avoid instances of abuse where creators are approached and solicited to create AIG-NCC:~\pquote{And so they’ll like DM that person and they’ll be like, `Hey, I’ll give you like forty bucks if you give me instructions on how to make a model based off someone'}~(\Pthirtyone{}).

Several communities also ban debates around politics and morality (\communityB{}, \communityC{}). Moderators have noted that this is to prevent instances of perceived political extremism such as \Pthirtytwo{} trying to prevent \pquote{[alt]-right and MAGA people coming to spew their hate}. 

However, given that \Pthirtyfour{} describes recent takedowns of AIG-NCC among offending websites as \pquote{all this sort of, you know, payment processor and, you know, political whatever}, these rules may also silence discussion of anti-abusive personal commitments through which anti-abusive norms could be established within the broader community.

\header{Age Attestation}
\communityA{} and \communityB{}, \edit{which focused on visual content creation,} also had public rules that by participating in the community, users self-attest to being at least 18 years of age.
Both communities included warnings that users may be banned if they are discovered to be minors.

\section{Enforcement and Governance Gaps (\ref{rq:enforcement})}
\label{subsec:enforcement-of-rules}

Three participants (\Pthree, \Pthirtyone, \Pthirtytwo) moderated communities and provided significant insight into the mechanisms of moderation within their communities, supplementing community members' narratives about moderator actions. 
While our discussion focuses on the three communities we recruited from, participants were often active on other Discord servers, where they observed varying levels of moderation.
For example, \Pthirtyone{} observed that on
\pquote{other servers and other parts of the internet [moderation is] done very poorly or not at all}, which they describe as an issue not unique to AIG-SC and, \pquote{just the reality of our internet.} 

\subsection{Enforcement}
\noindent\textbf{Community Owners and Moderators.}
Given the size of the communities and how quickly they may grow, the amount of content shared scales quickly. Moderation can be taxing \pquote{when you have that much activity with that many people, it’s very hard to keep track of everybody. So, that was the constant struggle}
(\Pthirtyone). They further added that because their community has a large number of members: \pquote{you just need people [moderating] 24/7, like you can’t go ten minutes, I would say, without viewing the server, without something being posted that needs to be checked.} 

Some chose to only allow output sharing in designated channels, which may help limit the number of channels that need to be more stringently monitored. \communityC{} leveraged Discord's spoiler tag feature, which blurs content by default, to reduce the visibility of shared content. Users must opt into viewing the content: \pquote{We do not shove our kinks at people...
it’s always behind spoilers... 
[members] have to click on it to see it. It’s their choice.}~(\Pseventeen)

\header{Community Members} 
\communityB{} and \communityC{} explicitly ask members to report others who are breaking the rules, particularly given the borderless nature of online communities. %
As \Pthirtyone{} describes,
\pquote{There’s like millions of celebrities...
maybe someone will recognize someone from  another country.} 
Proactive reports also help moderators stay informed as communities expand:
\begin{quote} \pquote{We did have a few instances where someone will...go around asking [for NCII]...
luckily our most active members...
they’ll be like, `Hey, we’re just letting you know that so-and-so sent me a DM saying they’ll pay me for basically deepfaking someone.' And so we’ll just ban the person once [the reporter] post[s] proof}~(\Pthirtyone).
\end{quote}

Community members often have vested interests in sustaining online spaces and safeguarding them (e.g., from deplatforming). \communityA{}  does not ask members to make reports, and perhaps relatedly, is the only community to note that users retain full liability for their actions. However, moderator \Pthree{} observed that in this community, non-moderator members enforce rules and norms \pquote{more seriously than I even
have... People see this as a valuable place, and they want it to be protected.}
Therefore, community members may proactively confront one another about behavior that violates stated rules. 
As described in Section~\ref{sec:rules:assets}, within \Psixteen{}'s text-based community, members will \pquote{gang up and go `get out of here'} if they see others asking for advice on creating content involving underage individuals.

\header{Self-Policing} 
As discussed in Section~\ref{sec:norms:commitments}, individual members may vary in their personal commitments regarding what content they will create, share, or view.
If members create content that does not violate their own personal commitments but are outside the bounds of community rules, they may refrain from sharing it to preserve the community: \pquote{If I can’t post it, I’ll just not post it... it’s less about keeping the community as clean as they want. It’s more about not getting banned on Discord} (\Peleven). 
In a few cases, participants were warned about their content and complied.
For example, \Pten{} refrained from further posting content because, \pquote{they told me that the character looks... underage. And I was like, `I’m making my best not to look, because the character is actually not underage'... 
[with] more cartoonish things, it’s very difficult to tell the age...
So, I stopped [posting].} %

\subsection{Governance Gaps}
\label{subsec:governance-gaps}

\noindent\textbf{Ambiguity in Enforcement.} 
AI-generated outputs introduce inherent ambiguity to moderation decision-making because, \pquote{when you generate something, it's not like you know their age because they don't exist. You just have to guess by looks and that can always be very, very subjective depending on where you're from in this world}~(\Pthirtyone{}). Ambiguity around age, particularly in animated content is also a challenge that platforms are contending with~\cite{petitLimitsZeroTolerance2025}.

Moderators often relied upon or partnered with community members to adjudicate such cases. As \Pthirtytwo{} explains,  
\pquote{it's a bit more complex to judge [images], so it's case by case... I also trust my community to judge if they think a piece of art represent a underage character.}
\Pthirtyone{} described how they not only moderated the appearance of the individual in an image, but its context and implications, \pquote{like school settings or school uniforms, anything that might suggest a younger age... 
we always are on the side of caution because we have no idea [what] the person [at the hosting platform] reviewing the server [will think].} In response to the same rule, \Pthirtyfour{} felt ``school settings'' was an imperfect barometer: \pquote{`Oh, your image had a chalkboard in the background?' No!... that just seems like kind of, overkill to me.}

\edit{Content that depicts fictional anime or cartoon characters is further complicated by different cultural norms.}
For example, \Pten{} explains that while they personally find underage anime characters to be \pquote{disgusting}, 
there are cultural and legal nuances to what content is permissible in different countries:
\pquote{In United States, there are some pages you can upload a full naked character and in Japan, you need to censor some parts of it, because of legislation regulations...There are other things that are clearly illegal in America that are not illegal in Japan\footnote{A 1999 Japanese law on child pornography which does not apply to AIG-CSAM~\cite{speedJapanStrugglesGray2025}.}... it’s very delicate.}

NCII also presents challenges for moderation since, \pquote{you don't know if this is purely AI-generated, or if it's a person. If it is a person, is it just a face swap, or is this their body, is this an original video, is than AI video. Do they even have consent? Yeah, we don’t know} (\Pthree). This challenge is further exacerbated by the rapid advancement of visual content generation techniques that make it difficult to distinguish AI-generated content from human-produced content.

\header{Legitimizing Enforcement: Laws and Terms of Service}
The AIG-SC communities our participants are involved in are typically created by a single individual or a small group.
In their early stages, communities tend to have fewer formalized rules and might be \pquote{fairly lenient because you’re not going to assume the worst of someone just because they’re acting up a bit}~(\Psixteen{}). Within smaller communities, it may also be easier to establish \textit{implicit norms} that do not need to be codified.
As membership grows, it becomes more difficult to monitor the volume of content. Growth can bring greater visibility to the community's activities as well, increasing potential legal risk for community owners.
In \Pthirtyone's experience, \pquote{We started off just a small Discord... but it kept growing... It was very lax in the beginning}. As they grew, they developed \pquote{strong rules} and \pquote{tend to be a little bit more strict... that's just out of safety and precaution for us.}

When creating and enforcing rules, moderators are motivated by their own commitments, their members, and by concerns around deplatforming---\pquote{it’s more about not getting banned on Discord}
(\Peleven)---as well as personal liability:~\pquote{%
if something happens... 
[Discord is not] going to look at the users, but the moderation team, and the person running the {[community]}}~({\Pthirtyone{}}). 
 
In an environment that prioritizes free expression, nonjudgment, and reducing stigma, several community members mentioned that it can be difficult to speak up about content that makes them uncomfortable, until it undeniably crosses a line (e.g., CSAM); even then, they may simply avoid it:
\pquote{I understand [it's others'] kink or whatever, but it's certainly not something that I'm into at all... 
that's something that over and over again I'm just X-ing out of, or hiding from my page} (\Pfortyseven).
Moderator \Pthree{} explains that if they overmoderate, their members will push back, saying 
\pquote{`I thought you were the uncensored place'... so I want to take the stance of legal[ality] is where I draw the line}.

Laws regarding CSAM offer a ``clear dividing line:'' 
\pquote{you can absolutely go to jail if you are spreading that content in a manner to encourage others to take part in, say, pedophilia. It’s very, very, very clear dividing line... it's very cut and dry} (\Pseventeen). 
\communityB{}~included warnings about complying with subpoenas and reporting to the U.S. National Center for Missing and Exploited Children in the United States, which operates a hotline for reporting child sexual abuse material~\cite{nationalcenterformissing&exploitedchildrenNationalCenterMissing2026}. 
Conversely, there are fewer established laws against creating NCII. 
Therefore, \Pthree{} found it harder to explain enforcement decisions demonstrating the importance of legal precedent in strengthening moderation decisions. For example, they have encountered cases where members \pquote{shared a video and they mentioned it's of their ex-girlfriend.} Although they had a discussion with the member,
they \pquote{don't know that there's a hard legal line on that, so I'm not even sure where to go.}

While laws may differ depending on the country, \Pthirtyone~feels that \pquote{mostly everyone who runs [the Discord server] is located either in the US or other Western countries and so we have to play by those rules at the end of the day.} Community members reside in any part of the world and may not agree since the \pquote{law on underage content can vary and can be legal as long as it's fictional in some country} (\Pthirtytwo). Therefore, \pquote{sometimes they’ll open a ticket or they’ll complain} (\Pthirtyone) but moderators can defend their decisions by instead pointing toward Discord's terms of service, which: \pquote{say no underage sexual depiction so I go with that, don't want my server to get nuked}~(\Pthirtytwo). In a different server, \Psixteen{} also noted that \pquote{the Discord terms of service would also tell us to kick people into underage content out... which is, again, fine in my view... go find some other people to bother.} The moderator of \communityA{} instead shared that beyond the bounds of the law in their jurisdiction, they, \pquote{don't even really care about Discord's 
rules...
I would rather make the community happy.}
In case their community is banned, \communityA{} has already created a backup Discord server and explains  \pquote{%
if we get banned, we have to go find somewhere else}~(\Pthree{}).

\section{Discussion}

Across our interviews, we observed a cultural ethos toward free knowledge exchange and uncensored AI generation. 
In part due to perceived rejection from other communities,
participants desired nonjudgmental spaces to discuss AIG-SC creation. 
While these cultural values enable creative and sexual exploration, they also create challenging tensions for moderation.
AIG-SC communities attract a range of individuals, including those who strongly oppose abusive content alongside some who use community resources to produce it.
Since AIG-SC communities can be surfaced through simple web searches and are at lower risk of being deplatformed, they are highly visible and accessible. Their governance, which attempts to prohibit abusive behavior while upholding their values, may be particularly important towards discouraging AIG-NCC offenses.
As such, we believe AIG-SC communities are important sites of intervention; we now provide several recommendations for doing so.

\header{Moderation Needs Assistance} 
Supporting sexual expression and condemning sexual violence are not contradictory actions; sex-positive education and perspectives can serve to mitigate actions and beliefs that perpetrate sexual violence~\cite{williams2013resolving, williams2015moving, 
peterson2024can,
sower2023kink, 
carmody2005ethical}, 
and help victim-survivors process  trauma~\cite{baggett2017sex, o2024think}.
However, clearly distinguishing what content is or is not abusive (particularly with little context) can be challenging and is an issue moderators face (Section~\ref{subsec:governance-gaps}). 
These moderation challenges mirror those encountered by platforms when attempting to prevent AIG-NCC generation at the model level, or the sharing of it on social platforms. Automated detectors of children in images have known inaccuracies~\cite{cretu2025evaluating} and distinguishing AIG-NCC from AIG-SC requires not nudity detection but rather identifying the likenesses of real people in generated content, which remains an unsolved problem with privacy implications. Beyond challenges in accurate content classification, community moderators report having to review an overwhelming amount of content, and handle the emotional toll caused by repeatedly viewing content flagged as AIG-NCC (Section~\ref{subsec:enforcement-of-rules}). Added to these domain-specific challenges are the general challenges faced by Discord moderators: that moderation relies on volunteer labor~\cite{matiasCivicLaborVolunteer2019} and encompasses text, voice, and video; the latter are more ephemeral and fundamentally more difficult to capture evidence for enforcement from~\cite{jiangModerationChallengesVoicebased2019}.

In light of these challenges, even members who hold strong personal commitments against AIG-NCC can begin to view this content as an inevitability (Section~\ref{subsec:socialization}, \textbf{\textit{Norm \#3}}). %
This minimizes harm as a byproduct of AI development and positions the technologies as neutral. 
This same type of justification is used by communities that condone AIG-NCC, such as commenters in face-swapping repositories: \pquote{like any technology, it can be used for good, or it can be abused}~\cite{newtonMyNSFWVideo2020}. 
 While non-abusive use cases for sexual and creative exploration exist~\cite{minkUnlimitedRealmExploration2026,riccio2024exposed,taylorUnStraighteningGenerativeAI2025}, complicating the narrative is the fact that those who create AIG-NCC have been known to co-opt these rationales~\cite{flynnSexualizedDeepfakeAbuse2025, hanCharacterizingMrDeepFakesSexual2025}, which can make it difficult to discern members' true motivations.

Given both the importance and difficulty of providing clear, justified, and correct moderation decisions in these communities in an AI industry that seeks to enable generation of ``NSFW stuff for your personal use... but not do stuff like make deepfakes''~\cite{altmansamasu/samaltmanandopenaiteamChatGPTHostingQA2024}, we recommend that future work investigate how to build sociotechnical processes that
work for moderators who lack corporate moderation resources.

\header{Prevention Focuses on Outputs Not Infrastructure}
In the AIG-SC communities we study, governance against AIG-NCC focuses nearly exclusively on outputs, not %
the assets and behaviors that enable those outputs (text-focused \communityC{} was the exception; Section~\ref{sec:rules:assets}). 
This is a significant governance gap, given that these communities serve as key hubs, amassing resources and exchanging knowledge to enable \textit{any} type of sexual content creation (Section~\ref{subsec:upskilling}). 

Data is collected and exchanged in accordance with the communities' norms of free knowledge (\textbf{\textit{Norm \#2}}); \communityA{} even goes so far to adopt practices to minimize possible legal repercussions. 
Several participants openly discussed non-consensually scraping and using data from adult websites and forums,
and
no participants discussed proactive measures to verify and remove content that could be illegal or violating (e.g., CSAM).
These are reminiscent of general ML practices, with CSAM discovered in LAION~\cite{thielGenerativeMLCSAM2023} and NudeNet~\cite{emanuelmaibergAIDatasetDetecting2025} datasets, and
NCII in nude-image datasets~\cite{princessacintaqiaStopNonconsensualUse2025}. 

Models are shared, bought, and traded without a clear understanding of what the model was trained to generate, or what it was trained on. In peer interactions, none of the participants mentioned vetting others in the community to ensure that they were not engaging in AIG-NCC prior to sharing advice; nor did those taking commissions discuss verifying authenticity from those claiming they were commissioning images depicting themselves.
Rather, the community\textit{ ``don't ask, don't tell''} (\textbf{\textit{Norm \#3}}) meant giving other members the benefit of the doubt and avoiding responsibility.
In combination with \textbf{\textit{Norm \#2}} of free information sharing, existing governance in %
AIG-SC communities can inadvertently enable AIG-NCC through shared knowledge, peer assistance, and shared technical assets (e.g., datasets, custom-created tools). 
Indeed, we find significant resemblance between the resource infrastructure supporting AIG-SC communities and those supporting AIG-NCC~\cite{medeirosCharacterizingResourceSharing2026,hanCharacterizingMrDeepFakesSexual2025}.

\header{Regulate AIG-NCC Creation, Commission, and Tech}
Although AIG-SC communities are frequented by individuals across the globe, regulation still plays a critical role in establishing a baseline for moderation and supporting enforcement decisions. 
While the impacts of regulation appear not to be uniform across AIG-SC spaces, particularly within AIG-NCII-supporting spaces such as 4chan~\cite{cuevasDeepfakePornographyResilient2026},
our findings suggest that among those who attempt to create AIG-SC in accordance with their own commitments and community norms, regulations provide impetus and justification for moderators to establish rules that echo regulation, justify the removal of problematic content, and reinforce norms
~(Section~\ref{sec:norms:moderation}). 
Legally influenced rules and their enforcement may be particularly important in communities.
First, it may be easier for impressionable newcomers to the space to justify joining such communities rather than more abuse-centric ones.
Second, moderators must be responsive to their community; if decisions are not appropriately communicated or justified, members may view these actions as at odds with held norms of uncensored creative freedom and open-mindedness (Section~\ref{sec:interactions-and-norms}, \textbf{\textit{Norm 1--3}}).
As members are not bound to any community, moderation perceived as improper may lead to migration towards less-regulated spaces that encourage abusive behavior.

Unfortunately, AIG-NCC regulation remains nascent and inconsistent across jurisdictions.
In the U.S., CSAM laws~\cite{fbiChildSexualAbuse2024} criminalize both the production and distribution of AIG-CSAM, while the TAKE IT DOWN Act~\cite{tedcruzS146TAKEIT2025} only criminalizes the nonconsensual distribution of AIG-NCII. 
Similarly, we find that posted community rules primarily focus on preventing CSAM outputs or, in one case, tools that enable CSAM generation (Section~\ref{subsect:community-rules}). 
Other regions have taken different approaches, including calls to ban the marketing of technology that enables AIG-NCC in the European Union~\cite{pieterhaeckEUSetBan2026}, criminalizing the viewing and possession of AIG-NCC in South Korea~\cite{gongSouthKoreaTackles2024, yimSouthKoreaCriminalise2024}, and using copyright law to ban any unauthorized use of an individual's likeness in Denmark~\cite{bryantDenmarkTackleDeepfakes2025}. 
However, these laws largely focus on the generated outputs themselves, without consideration for the broader ecosystem of assets that enable AIG-NCC creation. 
As one example, the UK recently amended an existing law to criminalize requests to commission AIG-NCII, regardless of whether creation takes place in the country. This accounts for offenders who may purposefully outsource creation to countries with little (or no) regulation~\cite{huiWhatKnowUK2026, parliamentoftheunitedkingdomDataUseAccess2025}. 
Without stringent regulation, offenders may believe 
there are no consequences to creating AIG-NCII, which helps 
rationalize their actions~\cite{flynnSexualizedDeepfakeAbuse2025}. 
Given our findings that regulation has a substantial downstream impact on platform policies, community governance, and the beliefs and actions of individual creators, we echo calls for regulation against the broader AIG-NCC enabling ecosystem.

\header{A Public Health Approach to AIG-NCC}
As scholars have previously identified, addressing sexual violence and child abuse requires a public health approach~\cite{mcmahon2000public, letourneauNeedComprehensivePublic2014} that invests in the development of interventions before harm occurs (\textit{primary prevention}) and immediately after harm has occurred (\textit{secondary prevention}). 
Despite their effectiveness at reducing overall offenses, such interventions are still limited for AIG-NCC.

\textit{Primary preventions} for AIG-NCC should consider the general populations that interact with these AI systems.
These could include developing and embedding computer science curricula that incorporate sex-positive education on consent proven to reduce offenses and offense-enabling beliefs~\cite{peterson2024can}. 
At a minimum, discussions on the ethics of non-consensual data collection and use, the dual use of generative AI, and the harms of AIG-NCC should be embedded.
This may help create and reinforce norms that reduce cyber-deviant behavior and crime~\cite {williams2006virtually}.

\textit{Primary and secondary interventions}
should more intentionally consider how to intervene with potential offenders (e.g., through visible behaviors).
While there is emerging work on deterrence messaging in the context of NCII~\cite{varunnagarajraoThinkTwiceYou}, deterrence messages for those seeking CSAM have already been integrated into online platforms (e.g., Pornhub, Google search), which early evidence suggests is
effective at reducing abuse-seeking searches ~\cite{scanlanRelationshipCSAMWarning2026, prichardEffectTherapeuticDeterrent2024}.
When messages are further combined with links to anonymous services, evidence suggests that a non-trivial number of (potential) offenders may stop and, in some cases, seek therapeutic help~\cite{scanlanRelationshipCSAMWarning2026, umbachDeterringSearchesChild2026}. 

Efforts are also needed to add friction to AIG-NCC creation. Experts should work to develop and evaluate effective peer-based deterrence messages to support moderation within AIG-SC communities. 
AI developers should consider implementing deterrence messages for prompts that appear to request, or seek advice on creating, AIG-NCC.
We urge that messages be developed in partnership with experts, as subtle variations can impact efficacy~\cite{ciardhaWarningMessageContent2026}.

Beyond single-turn deterrence messaging, comprehensive programs are needed to address AIG-NCC.
Educational programs shaping sexual violence-related attitudes and norms are recommended by the UN~\cite{unesco2018cse} and services such as peer support groups, self-help resources, anonymous helplines, and clinician-developed programs have been developed for CSAM-specific offenders and people at risk of offending~\cite{piche2018preventative, hillertWebBasedInitiativesPrevent2024}.
It is worth considering how similar services could be offered to reduce AIG-NCC.
It may also be useful to consider these services as alternatives to company-imposed punitive threats or deplatforming, which, %
may instead \textit{increase offenses} by moving those at-risk of offending to more extreme sects of the internet~\cite{cuevas2026deepfake}.

\section*{Acknowledgments}
This work was supported in part by NSF Award 2513313.
We thank Grace Brigham, Bernardo B. P. Medeiros, and Galen Weld for their detailed feedback on prior versions of this work. We thank Filipo Sharevski, Gianluca Stringhini, and Sharon Wang for providing advice on researcher and participant protection.

\bibliographystyle{ACM-Reference-Format}

\bibliography{references/bib-strings,references/polished-refs, references/digital-intimacy, references/jaron}

@string{sage = "{Sage Publications}"}

@string{routledge = "{Routledge}"}

@string{chi = "Proc. of {CHI}"}

@string{facct = "Proc. of {FAccT}"}

@string{usenix = "Proc. of {USENIX} Annual Technical Conference"}

@string{neurips = "Proc. of {NeurIPS}"}

@string{usenixsecurity = "Proc. of {USENIX} Security Symposium"}

@string{ccs = "Proc. of {CCS}"}

@string{soups = "Proc. of SOUPS"}

@string{cscw = "Proc. of {CSCW}"}

@article{newtonMyNSFWVideo2020,
	title = {My {NSFW} video has partial occlusion: deepfakes and the technological production of non-consensual pornography},
	volume = {7},
	issn = {2326-8743, 2326-8751},
	shorttitle = {My {NSFW} video has partial occlusion},
	url = {https://www.tandfonline.com/doi/full/10.1080/23268743.2019.1675091},
	doi = {10.1080/23268743.2019.1675091},
	language = {en},
	number = {4},
	urldate = {2025-08-19},
	journal = {Porn Studies},
	author = {Newton, Olivia B. and Stanfill, Mel},
	month = oct,
	year = {2020},
	pages = {398--414},
}

@misc{altmansamasu/samaltmanandopenaiteamChatGPTHostingQA2024,
	title = {r/{ChatGPT} is hosting a {Q}\&{A} with {OpenAI}’s {CEO} {Sam} {Altman}},
	url = {https://old.reddit.com/r/ChatGPT/comments/1coumbd/rchatgpt_is_hosting_a_qa_with_openais_ceo_sam/},
	author = {Altman, Sam (as u/samaltman) and OpenAI team},
	year = {2024},
}

@article{speedJapanStrugglesGray2025,
	title = {Japan struggles with gray zone of {AI} deepfakes exploiting children},
	url = {https://www.japantimes.co.jp/news/2025/04/27/japan/crime-legal/deepfake-ai-child-pornography/},
	urldate = {2025-09-11},
	journal = {The Japan Times},
	author = {Speed, Jessica},
	month = apr,
	year = {2025},
}

@misc{osf-aigsc-norms,
	title = {[{Additional} {Materials}] "{I} {Thought} {You} {Were} {The} {Uncensored} {Place}": {Norms}, {Rules}, and {Moderation} in {AI}-{Generated} {Sexual} {Content} {Communities}},
	url = {https://osf.io/g6a83/overview?view_only=e5acd9ae370c439f857d8ba33b6c51e1},
	author = {Qin, Lucy and Mink, Jaron and Redmiles, Elissa M.},
	month = aug,
	year = {2026},
}

@article{petitLimitsZeroTolerance2025,
	title = {The limits of ‘zero tolerance’ policies for animated pornographic media},
	issn = {2326-8743, 2326-8751},
	url = {https://www.tandfonline.com/doi/full/10.1080/23268743.2025.2491506},
	doi = {10.1080/23268743.2025.2491506},
	language = {en},
	urldate = {2026-01-23},
	journal = {Porn Studies},
	author = {Petit, Aurélie},
	month = may,
	year = {2025},
	pages = {1--18},
}

@misc{gongSouthKoreaTackles2024,
title = {South {Korea} investigates {Telegram} over alleged sexual deepfakes},
author = {Gong, Se Eun},
year = {2024},
month = sep,
day = {6},
howpublished = {NPR},
url = {https://www.npr.org/2024/09/06/nx-s1-5101891/south-korea-deepfake},
}

@misc{bryantDenmarkTackleDeepfakes2025,
	title = {Denmark to tackle deepfakes by giving people copyright to their own features},
	author = {Bryant, Miranda},
	month = jun,
	year = {2025},
	note = {Published: The Guardian},
}

@misc{yimSouthKoreaCriminalise2024,
	title = {South {Korea} to criminalise watching or possessing sexually explicit deepfakes},
	author = {Yim, Hyunsu},
	month = sep,
	year = {2024},
	note = {Published: Reuters},
}

@inproceedings{juliakiesermanCaughtMafiaRomance2026,
	title = {Caught in a {Mafia} {Romance}: {How} {Users} {Explore} {Intimate} {Narratives} with {Chatbots}},
	booktitle = {Proceedings of the 2026 {CHI} {Conference} on {Human} {Factors} in {Computing} {Systems}},
	publisher = {Association for Computing Machinery},
  author={Kieserman, Julia and Mai, Cat and Lignell, Sara and Qin, Lucy and Andreou, Athanasios and McCoy, Damon and Bellini, Rosanna},
	year = {2026},
}

@article{emanuelmaibergAIDatasetDetecting2025,
	title = {{AI} {Dataset} for {Detecting} {Nudity} {Contained} {Child} {Sexual} {Abuse} {Images}},
	url = {https://www.404media.co/ai-dataset-for-detecting-nudity-contained-child-sexual-abuse-images/},
	journal = {404 Media},
	author = {Emanuel Maiberg},
	month = oct,
	year = {2025},
}

@article{lapointeGovernanceAIgeneratedPornography2026,
	title = {The governance of {AI}-generated pornography platforms: {A} content analysis},
	issn = {1461-4448, 1461-7315},
	shorttitle = {The governance of {AI}-generated pornography platforms},
	url = {https://journals.sagepub.com/doi/10.1177/14614448261421873},
	doi = {10.1177/14614448261421873},
	language = {en},
	urldate = {2026-03-20},
	journal = {New Media \& Society},
	author = {Lapointe, Valerie A and Dubé, Simon and Petit, Aurélie and Kessai, Tinhinane and Rukhlyadyev, Sophia and Gravel, Vivianne and Lafortune, David},
	month = feb,
	year = {2026},
	pages = {14614448261421873},
}

@misc{cuevasDeepfakePornographyResilient2026,
	title = {Deepfake {Pornography} is {Resilient} to {Regulatory} and {Platform} {Shocks}},
	url = {http://arxiv.org/abs/2602.02754},
	doi = {10.48550/arXiv.2602.02754},
	urldate = {2026-04-23},
	publisher = {arXiv},
	author = {Cuevas, Alejandro and Ribeiro, Manoel Horta},
	month = feb,
	year = {2026},
	note = {arXiv:2602.02754 [cs]},
}

@misc{brighamExaminingRisksAI2026,
	title = {Examining {Risks} in the {AI} {Companion} {Application} {Ecosystem}},
	url = {http://arxiv.org/abs/2603.13620},
	doi = {10.48550/arXiv.2603.13620},
	urldate = {2026-04-24},
	publisher = {arXiv},
	author = {Brigham, Natalie Grace and Qin, Lucy and Kohno, Tadayoshi},
	month = mar,
	year = {2026},
	note = {arXiv:2603.13620 [cs]},
}

@article{jiangModerationChallengesVoicebased2019,
	title = {Moderation {Challenges} in {Voice}-based {Online} {Communities} on {Discord}},
	volume = {3},
	issn = {2573-0142},
	url = {https://dl.acm.org/doi/10.1145/3359157},
	doi = {10.1145/3359157},
	language = {en},
	number = {CSCW},
	urldate = {2026-04-24},
	journal = {Proceedings of the ACM on Human-Computer Interaction},
	author = {Jiang, Jialun Aaron and Kiene, Charles and Middler, Skyler and Brubaker, Jed R. and Fiesler, Casey},
	month = nov,
	year = {2019},
	pages = {1--23},
}

@article{umbachAIgeneratedImagebasedSexual2026,
	title = {{AI}-generated image-based sexual abuse: {Perpetration} and consumption across three regions},
	volume = {179},
	issn = {07475632},
	shorttitle = {{AI}-generated image-based sexual abuse},
	url = {https://linkinghub.elsevier.com/retrieve/pii/S0747563226000324},
	doi = {10.1016/j.chb.2026.108935},
	language = {en},
	urldate = {2026-04-24},
	journal = {Computers in Human Behavior},
	author = {Umbach, Rebecca and Henry, Nicola and Shelby, Renee and Stevens, Gemma and Gonzalez-Pons, Kwynn},
	month = jun,
	year = {2026},
	pages = {108935},
}

@article{matiasCivicLaborVolunteer2019,
	title = {The {Civic} {Labor} of {Volunteer} {Moderators} {Online}},
	volume = {5},
	issn = {2056-3051, 2056-3051},
	url = {https://journals.sagepub.com/doi/10.1177/2056305119836778},
	doi = {10.1177/2056305119836778},
	language = {en},
	number = {2},
	urldate = {2026-04-24},
	journal = {Social Media + Society},
	author = {Matias, J. Nathan},
	month = apr,
	year = {2019},
	pages = {2056305119836778},
}

@misc{medeirosCharacterizingResourceSharing2026,
	title = {Characterizing {Resource} {Sharing} {Practices} on {Underground} {Internet} {Forum} {Synthetic} {Non}-{Consensual} {Intimate} {Image} {Content} {Creation} {Communities}},
	copyright = {arXiv.org perpetual, non-exclusive license},
	url = {https://arxiv.org/abs/2604.12190},
	doi = {10.48550/ARXIV.2604.12190},
	urldate = {2026-04-27},
	publisher = {arXiv},
	author = {Medeiros, Bernardo B. P. and Jadhav, Malvika and Lu, Allison and Kohno, Tadayoshi and Bindschaedler, Vincent and Butler, Kevin R. B.},
	year = {2026},
	note = {Version Number: 1},
}

@article{chanMusksGrokChatbot2026,
	title = {Musk's {Grok} chatbot faces {EU} privacy investigation over sexualized deepfake images},
	journal = {PBS},
	author = {Chan, Kelvin},
	month = feb,
	year = {2026},
}

@article{wilsonHundredsNonconsensualAI2026,
	title = {Hundreds of nonconsensual {AI} images being created by {Grok} on {X}, data shows},
	url = {https://www.theguardian.com/technology/2026/jan/08/grok-x-nonconsensual-images},
	journal = {The Guardian},
	author = {Wilson, Jason},
	month = jan,
	year = {2026},
}

@inproceedings{zengHiddenDesiresExploring2026,
	address = {Barcelona Spain},
	title = {The {Hidden} {Desires}: {Exploring} {Chinese} {Women}'s {Erotic} {Role}-{Play} with {AI} {Chatbots}},
	isbn = {979-8-4007-2278-3},
	shorttitle = {The {Hidden} {Desires}},
	url = {https://dl.acm.org/doi/10.1145/3772318.3791580},
	doi = {10.1145/3772318.3791580},
	language = {en},
	urldate = {2026-04-29},
	booktitle = {Proceedings of the 2026 {CHI} {Conference} on {Human} {Factors} in {Computing} {Systems}},
	publisher = {ACM},
	author = {Zeng, Jiayi and Wu, Mengmeng and Lan, Xingyu},
	month = apr,
	year = {2026},
	pages = {1--15},
}

@misc{tsuchiyaLargeScaleStudyTelegram2026,
	title = {A {Large}-{Scale} {Study} of {Telegram} {Bots}},
	url = {http://arxiv.org/abs/2603.24302},
	doi = {10.48550/arXiv.2603.24302},
	urldate = {2026-05-06},
	publisher = {arXiv},
	author = {Tsuchiya, Taro and Yu, Haoxiang and Marjanov, Tina and Hutchings, Alice and Christin, Nicolas and Cuevas, Alejandro},
	month = apr,
	year = {2026},
	note = {arXiv:2603.24302 [cs]},
}

@misc{umbachDeterringSearchesChild2026,
	title = {Deterring {Searches} for {Child} {Sexual} {Abuse} {Material} on {Google} {Search} and {Promoting} {Help}-{Seeking}},
	copyright = {Creative Commons Attribution 4.0 International},
	url = {https://arxiv.org/abs/2606.06126},
	doi = {10.48550/ARXIV.2606.06126},
	urldate = {2026-08-11},
	publisher = {arXiv},
	author = {Umbach, Rebecca and Hunt, Griffin and Buckley, John and Scanlan, Joel and Ciardha, Caoilte \'{O} and Quayle, Ethel and Heasman, Ainslie and von Heyden, Maximilian and Letourneau, Elizabeth and Findlater, Donald and Insoll, Tegan and Wortley, Richard and Steel, Chad and Roy, Abhishek},
	year = {2026},
	note = {Version Number: 2},
}

@misc{ciardhaWarningMessageContent2026,
	title = {Warning {Message} {Content} {Increases} {Help} {Seeking} in a {Large}-{Scale} {Dark} {Web} {CSAM} {Intervention}},
	copyright = {Creative Commons Attribution 4.0 International},
	url = {https://arxiv.org/abs/2606.06417},
	doi = {10.48550/ARXIV.2606.06417},
	urldate = {2026-08-11},
	publisher = {arXiv},
	author = {Ciardha, Caoilte \'{O} and Scanlan, Joel and Insoll, Tegan and Nurmi, Juha and Vaaranen-Valkonen, Nina},
	year = {2026},
	note = {Version Number: 2},
}

@techreport{varunnagarajraoThinkTwiceYou,
	title = {Think {Twice} {Before} {You} {Search}: {Deterrence} {Messaging} {Designs} to {Prevent} {Searches} for {Non}-{Consensual} {Intimate} {Images}},
	url = {https://cdt.org/insights/think-twice-before-you-search-deterrence-messaging-designs- to-prevent-searches-for-non-consensual-intimate-images},
	institution = {Center for Democracy \& Technology},
	author = {{Varun Nagaraj Rao} and {Dhanaraj Thakur}},
	month = nov,
	year = {2025},
}

@article{prichardEffectTherapeuticDeterrent2024,
	title = {The effect of therapeutic and deterrent messages on {Internet} users attempting to access ‘barely legal’ pornography},
	volume = {155},
	issn = {01452134},
	url = {https://linkinghub.elsevier.com/retrieve/pii/S0145213424003454},
	doi = {10.1016/j.chiabu.2024.106955},
	language = {en},
	urldate = {2026-08-11},
	journal = {Child Abuse \& Neglect},
	author = {Prichard, Jeremy and Wortley, Richard and Watters, Paul and Spiranovic, Caroline and Scanlan, Joel},
	month = sep,
	year = {2024},
	pages = {106955},
}

@article{scanlanRelationshipCSAMWarning2026,
	title = {The {Relationship} {Between} a {CSAM} {Warning} {Messaging} {Chatbot} and {User} {Behavior} on {Pornhub}},
	volume = {21},
	issn = {1556-4886, 1556-4991},
	url = {https://www.tandfonline.com/doi/full/10.1080/15564886.2025.2606857},
	doi = {10.1080/15564886.2025.2606857},
	language = {en},
	number = {3},
	urldate = {2026-08-11},
	journal = {Victims \& Offenders},
	author = {Scanlan, Joel and Hall, Lauren C. and Watters, Paul A. and Wortley, Richard and Prichard, Jeremy},
	month = apr,
	year = {2026},
	pages = {574--602},
}

@article{hillertWebBasedInitiativesPrevent2024,
	title = {Web-{Based} {Initiatives} to {Prevent} {Sexual} {Offense} {Perpetration}: {A} {Systematic} {Review}},
	volume = {26},
	issn = {1523-3812, 1535-1645},
	shorttitle = {Web-{Based} {Initiatives} to {Prevent} {Sexual} {Offense} {Perpetration}},
	url = {https://link.springer.com/10.1007/s11920-024-01489-1},
	doi = {10.1007/s11920-024-01489-1},
	language = {en},
	number = {4},
	urldate = {2026-08-11},
	journal = {Current Psychiatry Reports},
	author = {Hillert, Jana and Haubrock, Lina Sophie and Dekker, Arne and Briken, Peer},
	month = apr,
	year = {2024},
	pages = {121--133},
}

@techreport{thielGenerativeMLCSAM2023,
	title = {Generative {ML} and {CSAM}: {Implications} and {Mitigations}},
	copyright = {Creative Commons Attribution Non Commercial No Derivatives 4.0 International},
	shorttitle = {Generative {ML} and {CSAM}},
	url = {https://purl.stanford.edu/jv206yg3793},
	doi = {10.25740/JV206YG3793},
	urldate = {2026-08-11},
	institution = {Stanford Digital Repository},
	author = {Thiel, David and Stroebel, Melissa and Portnoff, Rebecca and Hancock, Jeffrey and Scyphers, Cassandra and O'Gorman, Tim and DiResta, Renée},
	year = {2023},
}

@article{letourneauNeedComprehensivePublic2014,
	title = {The {Need} for a {Comprehensive} {Public} {Health} {Approach} to {Preventing} {Child} {Sexual} {Abuse}},
	volume = {129},
	issn = {0033-3549, 1468-2877},
	url = {https://journals.sagepub.com/doi/10.1177/003335491412900303},
	doi = {10.1177/003335491412900303},
	language = {en},
	number = {3},
	urldate = {2026-08-13},
	journal = {Public Health Reports®},
	author = {Letourneau, Elizabeth J. and Eaton, William W. and Bass, Judith and Berlin, Frederick S. and Moore, Stephen G.},
	month = may,
	year = {2014},
	pages = {222--228},
}

@misc{missingkidsHome,
  author       = {{National Center for Missing \& Exploited Children}},
  title        = {MissingKids.org: Home},
  year         = {2026},
  howpublished = {\url{https://www.missingkids.org/home}},
  note         = {Accessed: 2026-01-14},
  organization = {National Center for Missing \& Exploited Children},
}

@article{10.1145/3415226,
author = {Dym, Brianna and Fiesler, Casey},
title = {Social Norm Vulnerability and its Consequences for Privacy and Safety in an Online Community},
year = {2020},
issue_date = {October 2020},
publisher = {Association for Computing Machinery},
address = {New York, NY, USA},
volume = {4},
number = {CSCW2},
url = {https://doi.org/10.1145/3415226},
doi = {10.1145/3415226},
journal = {Proc. ACM Hum.-Comput. Interact.},
month = oct,
articleno = {155},
numpages = {24}
}

@book{siitonen2007social,
  title={Social interaction in online multiplayer communities},
  author={Siitonen, Marko},
  number={74},
  year={2007},
  publisher={University of Jyv{\"a}skyl{\"a}}
}

@article{opp2001norms,
  title={How do norms emerge? An outline of a theory},
  author={Opp, Karl-Dieter},
  journal={Mind \& Society},
  volume={2},
  number={1},
  pages={101--128},
  year={2001},
  publisher={Springer}
}

@article{liefbroer2010bringing,
  title={Bringing norms back in: A theoretical and empirical discussion of their importance for understanding demographic behaviour},
  author={Liefbroer, Aart C and Billari, Francesco C},
  journal={Population, space and place},
  volume={16},
  number={4},
  pages={287--305},
  year={2010},
  publisher={Wiley Online Library}
}

@article{rashidi2020s,
  title={"It's easier than causing confrontation": sanctioning strategies to maintain social norms and privacy on social media},
  author={Rashidi, Yasmeen and Kapadia, Apu and Nippert-Eng, Christena and Su, Norman Makoto},
  journal={Proceedings of the ACM on human-computer interaction},
  volume={4},
  number={CSCW1},
  pages={1--25},
  year={2020},
  publisher={ACM New York, NY, USA}
}

@article{o2024think,
  title={“I Think I Just like Having Sex”: A Qualitative Study of Sexual Assault Survivors and Their Sexual Pleasure},
  author={O’Callaghan, Erin and Lorenz, Katherine},
  journal={Sex Roles},
  volume={90},
  number={9},
  pages={1169--1187},
  year={2024},
  publisher={Springer}
}

@article{baggett2017sex,
  title={Sex-positive assessment and treatment among female trauma survivors},
  author={Baggett, Linda R and Eisen, Ethan and Gonzalez-Rivas, Sara and Olson, Lacy A and Cameron, Rebecca P and Mona, Linda R},
  journal={Journal of clinical psychology},
  volume={73},
  number={8},
  pages={965--974},
  year={2017},
  publisher={Wiley Online Library}
}

@article{cowburn2005confidentiality,
  title={Confidentiality and public protection: Ethical dilemmas in qualitative research with adult male sex offenders},
  author={Cowburn, Malcolm},
  journal={Journal of sexual aggression},
  volume={11},
  number={1},
  pages={49--63},
  year={2005},
  publisher={Taylor \& Francis}
}

@article{williams2013resolving,
  title={Resolving social problems associated with sexuality: Can a ``sex-positive''' approach help?},
  author={Williams, DJ and Prior, Emily and Wegner, Jenna},
  journal={Social work},
  volume={58},
  number={3},
  pages={273--276},
  year={2013},
  publisher={Oxford University Press}
}

@article{sen1977rational,
  title={Rational fools: A critique of the behavioral foundations of economic theory},
  author={Sen, Amartya K},
  journal={Philosophy \& public affairs},
  pages={317--344},
  year={1977},
  publisher={JSTOR}
}

@article{ostrom2011background,
  title={Background on the institutional analysis and development framework},
  author={Ostrom, Elinor},
  journal={Policy studies journal},
  volume={39},
  number={1},
  pages={7--27},
  year={2011},
  publisher={Wiley Online Library}
}

@misc{maiberg25_civitai_cutoff,
  author       = "Emanuel Maiberg",
  title        = "{Civitai, Site Used to Generate AI Porn, Cut Off by Credit Card Processor}",
  note         = {\url{https://www.404media.co/civitai-site-used-to-generate-ai-porn-cut-off-by-credit-card-processor/}},
  year         = "2025",
  howpublished = "404 Media (online)",
  month        = may,
}

@article{wilson2002effects,
  title={Effects of interviewer characteristics on reported sexual behavior of California Latino couples},
  author={Wilson, Sandra R and Brown, Nancy L and Mejia, Carolina and Lavori, Philip W},
  journal={Hispanic Journal of Behavioral Sciences},
  volume={24},
  number={1},
  pages={38--62},
  year={2002},
  publisher={Sage Publications Sage CA: Thousand Oaks, CA}
}

@article{harris2019joining,
  title={Joining together online: the trajectory of CSCW scholarship on group formation},
  author={Harris, Alexa M and G{\'o}mez-Zar{\'a}, Diego and DeChurch, Leslie A and Contractor, Noshir S},
  journal={Proceedings of the ACM on Human-Computer Interaction},
  volume={3},
  number={CSCW},
  pages={1--27},
  year={2019},
  publisher={ACM New York, NY, USA}
}

@misc{associatedpress2026_elonSuesMN,
  author       = {Marc Levy and Barbara Ortutay},
  title        = {Elon Musk’s {xAI} sues Minnesota over its first-in-the-nation law banning ‘nudification’ technology},
  note         = {\url{https://apnews.com/article/minnesota-artificial-intelligence-nudification-x-elon-musk-deepfake-131184be939d540de093b567b12c9e16}},
  year         = "2026",
  organization = "Associated Press",
}

@inproceedings{mink2022Deepphish,
  title={$\{$DeepPhish$\}$: Understanding user trust towards artificially generated profiles in online social networks},
  author={Mink, Jaron and Luo, Licheng and Barbosa, Nat{\~a} M and Figueira, Olivia and Wang, Yang and Wang, Gang},
  booktitle={31st USENIX Security Symposium (USENIX Security 22)},
  pages={1669--1686},
  year={2022}
}

@inproceedings{mink2024Moderation,
author = {Mink, Jaron and Wei, Miranda and Munyendo, Collins W. and Hugenberg, Kurt and Kohno, Tadayoshi and Redmiles, Elissa M. and Wang, Gang},
title = {It's Trying Too Hard To Look Real: Deepfake Moderation Mistakes and {Identity-Based} Bias},
year = {2024},
isbn = {9798400703300},
publisher = {Association for Computing Machinery},
address = {New York, NY, USA},
doi = {10.1145/3613904.3641999},
booktitle = {Proceedings of the 2024 CHI Conference on Human Factors in Computing Systems},
articleno = {778},
numpages = {20},
location = {Honolulu, HI, USA},
series = {CHI '24}
}

@inproceedings{pendse2025testing,
  title={When Testing {AI} Tests Us: Safeguarding Mental Health on the Digital Frontlines},
  author={Pendse, Sachin R and Gergle, Darren and Kornfield, Rachel and Meyerhoff, Jonah and Mohr, David and Suh, Jina and Wescott, Annie and Williams, Casey and Schleider, Jessica},
  booktitle={Proceedings of the 2025 ACM Conference on Fairness, Accountability, and Transparency},
  pages={1793--1804},
  year={2025}
}

@misc{EFF_Flawed_Take_It_Down_Act_2025,
  author       = {Joe Mullin},
  title        = {The {TAKE IT DOWN} Act: A Flawed Attempt to Protect Victims That Will Lead to Censorship},
  year         = {2025},
  month        = {February},
  day          = {11},
  howpublished = {Electronic Frontier Foundation},
  note         = {\url{https://www.eff.org/deeplinks/2025/02/take-it-down-act-flawed-attempt-protect-victims-will-lead-censorship}} 
}

@article{doerfler2021m,
  title={" I'm a Professor, which isn't usually a dangerous job": Internet-facilitated Harassment and Its Impact on Researchers},
  author={Doerfler, Periwinkle and Forte, Andrea and De Cristofaro, Emiliano and Stringhini, Gianluca and Blackburn, Jeremy and McCoy, Damon},
  journal={Proceedings of the ACM on Human-Computer Interaction},
  volume={5},
  number={CSCW2},
  pages={1--32},
  year={2021},
  publisher={ACM New York, NY, USA}
}

@article{schopke2024volunteer,
  title={Why do volunteer content moderators quit? Burnout, conflict, and harmful behaviors},
  author={Sch{\"o}pke-Gonzalez, Angela M and Atreja, Shubham and Shin, Han Na and Ahmed, Najmin and Hemphill, Libby},
  journal={New Media \& Society},
  volume={26},
  number={10},
  pages={5677--5701},
  year={2024},
  publisher={SAGE Publications Sage UK: London, England}
}

@inproceedings{wohn2019volunteer,
  title={Volunteer moderators in twitch micro communities: How they get involved, the roles they play, and the emotional labor they experience},
  author={Wohn, Donghee Yvette},
  booktitle={Proceedings of the 2019 CHI conference on human factors in computing systems},
  pages={1--13},
  year={2019}
}

@article{lloyd2025there,
  title={'There Has To Be a Lot That We're Missing': Moderating {AI-Generated} Content on Reddit},
  author={Lloyd, Travis and Reagle, Joseph and Naaman, Mor},
  journal={Proceedings of the ACM on Human-Computer Interaction},
  volume={9},
  number={7},
  pages={1--24},
  year={2025},
  publisher={ACM New York, NY, USA}
}

@inproceedings{li2022all,
  title={All that’s happening behind the scenes: Putting the spotlight on volunteer moderator labor in reddit},
  author={Li, Hanlin and Hecht, Brent and Chancellor, Stevie},
  booktitle={Proceedings of the International AAAI Conference on Web and Social Media},
  volume={16},
  pages={584--595},
  year={2022}
}

@misc{unesco2018cse,
  author       = {{UNESCO}},
  title        = {Comprehensive Sexuality Education to Prevent Gender-Based Violence},
  year         = {2018},
  month        = mar,
  day          = {13},
  howpublished = {UNESCO},
  note         = {Last updated April 20, 2023},
  url          = {https://www.unesco.org/en/articles/comprehensive-sexuality-education-prevent-gender-based-violence}
}

@article{cui2026celebrities,
  title={From Celebrities to Anyone: Characterizing AI Nudification Content, Technology, and Community Dynamics on 4chan},
  author={Cui, Chi and Wu, Yixin and Zhang, Yang},
  journal={arXiv preprint arXiv:2606.27234},
  year={2026}
}

@article{daffalla2026dual,
  title={Dual-Use AI Face Swap Apps Are Mostly Unsafe: A Systematic Safety Audit},
  author={Daffalla, Alaa and Chao, Sarah and Zeng, Eric},
  journal={arXiv preprint arXiv:2605.24735},
  year={2026}
}

@article{cuevas2026deepfake,
  title={Deepfake Pornography is Resilient to Regulatory and Platform Shocks},
  author={Cuevas, Alejandro and Ribeiro, Manoel Horta},
  journal={arXiv preprint arXiv:2602.02754},
  year={2026}
}

@book{williams2006virtually,
  title={Virtually criminal: Crime, deviance and regulation online},
  author={Williams, Matthew},
  year={2006},
  publisher={Routledge}
}

@article{mcmahon2000public,
  title={The public health approach to the prevention of sexual violence},
  author={McMahon, Pamela M},
  journal={Sexual Abuse: A Journal of Research and Treatment},
  volume={12},
  number={1},
  pages={27--36},
  year={2000},
  publisher={Springer}
}

@article{piche2018preventative,
  title={Preventative services for sexual offenders},
  author={Pich{\'e}, Lyne and Mathesius, Jeffrey and Lussier, Patrick and Schweighofer, Anton},
  journal={Sexual Abuse},
  volume={30},
  number={1},
  pages={63--81},
  year={2018},
  publisher={Sage Publications Sage CA: Los Angeles, CA}
}

@inproceedings{10.1145/3772318.3790415,
author = {Gao, Lan and Ahmed, Abani and Chen, Oscar and Reyl, Margaux and Cheema, Zayna and Feamster, Nick and Tan, Chenhao and Thomas, Kurt and Chetty, Marshini},
title = {Governance of AI-Generated Content: A Case Study on Social Media Platforms},
year = {2026},
isbn = {9798400722783},
publisher = {Association for Computing Machinery},
address = {New York, NY, USA},
url = {https://doi.org/10.1145/3772318.3790415},
doi = {10.1145/3772318.3790415},
booktitle = {Proceedings of the 2026 CHI Conference on Human Factors in Computing Systems},
articleno = {1604},
numpages = {22},
location = {
},
series = {CHI '26}
}

@inproceedings{mayworm2024misgendered,
  title={Misgendered during moderation: How transgender bodies make visible cisnormative content moderation policies and enforcement in a meta oversight board case},
  author={Mayworm, Samuel and Albert, Kendra and Haimson, Oliver L},
  booktitle={Proceedings of the 2024 ACM conference on fairness, accountability, and transparency},
  pages={301--312},
  year={2024}
}

@article{blunt2021automating,
  title={Automating whorephobia: sex, technology and the violence of deplatforming: an interview with hacking//hustling},
  author={Blunt, Danielle and Stardust, Zahra},
  journal={Porn Studies},
  volume={8},
  number={4},
  pages={350--366},
  year={2021},
  publisher={Taylor \& Francis}
}

@article{haimson2021disproportionate,
  title={Disproportionate removals and differing content moderation experiences for conservative, transgender, and black social media users: Marginalization and moderation gray areas},
  author={Haimson, Oliver L and Delmonaco, Daniel and Nie, Peipei and Wegner, Andrea},
  journal={Proceedings of the ACM on Human-computer Interaction},
  volume={5},
  number={CSCW2},
  pages={1--35},
  year={2021},
  publisher={ACM New York, NY, USA}
}

@inproceedings{taylor2025straightening,
  title={Un-Straightening Generative AI: How Queer Artists Surface and Challenge Model Normativity},
  author={Taylor, Jordan and Mire, Joel and Spektor, Franchesca and DeVrio, Alicia and Sap, Maarten and Zhu, Haiyi and Fox, Sarah E},
  booktitle={Proceedings of the 2025 ACM Conference on Fairness, Accountability, and Transparency},
  pages={951--963},
  year={2025}
}

@article{zolides2021gender,
  title={Gender moderation and moderating gender: Sexual content policies in Twitch’s community guidelines},
  author={Zolides, Andrew},
  journal={New Media \& Society},
  volume={23},
  number={10},
  pages={2999--3015},
  year={2021},
  publisher={Sage Publications Sage UK: London, England}
}

@article{tiidenberg2021sex,
  title={Sex, power and platform governance},
  author={Tiidenberg, Katrin},
  journal={Porn Studies},
  volume={8},
  number={4},
  pages={381--393},
  year={2021},
  publisher={Taylor \& Francis}
}

@techreport{chandraReducingRisksPosed2024,
  author = {Bilva Chandra and Jesse Dunietz and Kathleen Roberts and Yooyoung Lee and Peter Fontana and George Awad},
  title = {Reducing Risks Posed by Synthetic Content An Overview of Technical Approaches to Digital Content Transparency},
  year = {2024},
  institution = {NIST Trustworthy and Responsible AI},
  language = {en},
}

@misc{fbiChildSexualAbuse2024,
	title = {Child {Sexual} {Abuse} {Material} {Created} by {Generative} {AI} and {Similar} {Online} {Tools} is {Illegal}},
	author = {{FBI}},
	year = {2024},
    url = {{https://www.ic3.gov/PSA/2024/PSA240329}}
}

@misc{emanuelmaibergBoomingAIPimping2024,
	title = {Inside the {Booming} '{AI} {Pimping}' {Industry}},
	url = {https://www.404media.co/inside-the-booming-ai-pimping-industry-3/},
	howpublished = {404 Media},
	author = {Emanuel Maiberg and Jason Koebler},
	month = nov,
	year = {2024}
}

@misc{emanuelmaibergAIPornMarketplace2023,
	title = {Inside the {AI} {Porn} {Marketplace} {Where} {Everything} and {Everyone} {Is} for {Sale}},
	url = {https://www.404media.co/inside-the-ai-porn-marketplace-where-everything-and-everyone-is-for-sale/},
	howpublished = {404 Media},
	author = {Emanuel Maiberg},
	month = aug,
	year = {2023},
}

@misc{maibergCivitaiBanReal2025,
	title = {Civitai {Ban} of {Real} {People} {Content} {Deals} {Major} {Blow} to the {Nonconsensual} {AI} {Porn} {Ecosystem}},
	howpublished = {404 Media},
	author = {Emanuel Maiberg},
    url = {https://www.404media.co/civitai-ban-of-real-people-content-deals-major-blow-to-the-nonconsensual-ai-porn-ecosystem/},
	year = {2025},
}

@misc{tedcruzS146TAKEIT2025,
	title = {S.146 - {TAKE} {IT} {DOWN} {Act}},
	url = {https://www.congress.gov/bill/119th-congress/senate-bill/146},
	author = {{Ted Cruz}},
	month = may,
	year = {2025},
}

@misc{minkUnlimitedRealmExploration2026,
	title = {{"Unlimited Realm of Exploration and Experimentation"}: {Methods} and {Motivations} of {AI-Generated} {Sexual} {Content} {Creators}},
	author = {Mink, Jaron and Qin, Lucy and Redmiles, Elissa M.},
	year = {2026},
    booktitle = facct,
}

@inproceedings{gibsonAnalyzingAINudification2025,
	title = {Analyzing the \{{AI}\} {Nudification} {Application} {Ecosystem}},
	booktitle = usenixsecurity,
	author = {Gibson, Cassidy and Olszewski, Daniel and Brigham, Natalie Grace and Crowder, Anna and Butler, Kevin RB and Traynor, Patrick and Redmiles, Elissa M. and Kohno, Tadayoshi},
	year = {2025},
}

@inproceedings{hawkinsDeepfakesDemandRise2025,
	title = {Deepfakes on {Demand}: {The} rise of accessible non-consensual deepfake image generators},	
	booktitle = facct,
	author = {Hawkins, Will and Mittelstadt, Brent and Russell, Chris},
	year = {2025},
}

@inproceedings{brighamViolationMyBody2024,
	title = {{"{Violation} of my \{body:\}" {Perceptions} of \{{AI}-generated\} non-consensual (intimate) imagery}},
	booktitle = soups,
	author = {Brigham, Natalie Grace and Wei, Miranda and Kohno, Tadayoshi and Redmiles, Elissa M.},
	year = {2024},
}

@inproceedings{baSurrogatepromptBypassingSafety2024,
	title = {Surrogateprompt: {Bypassing} the safety filter of text-to-image models via substitution},
	booktitle = ccs,
	author = {Ba, Zhongjie and Zhong, Jieming and Lei, Jiachen and Cheng, Peng and Wang, Qinglong and Qin, Zhan and Wang, Zhibo and Ren, Kui},
	year = {2024},
}

@inproceedings{taylorUnStraighteningGenerativeAI2025,
	title = {Un-{Straightening} {Generative} {AI}: {How} {Queer} {Artists} {Surface} and {Challenge} {Model} {Normativity}},
	booktitle = facct,
	author = {Taylor, Jordan and Mire, Joel and Spektor, Franchesca and DeVrio, Alicia and Sap, Maarten and Zhu, Haiyi and Fox, Sarah E},
	year = {2025},
}

@inproceedings{hanCharacterizingMrDeepFakesSexual2025,
	title = {Characterizing the \{{MrDeepFakes}\} {Sexual} {Deepfake} {Marketplace}},
	booktitle = usenixsecurity,
	author = {Han, Catherine and Li, Anne and Kumar, Deepak and Durumeric, Zakir},
	year = {2025},
}

@inproceedings{umbachNonconsensualSyntheticIntimate2024,
	title = {Non-consensual synthetic intimate imagery: {Prevalence}, attitudes, and knowledge in 10 countries},
	booktitle = chi,
	author = {Umbach, Rebecca and Henry, Nicola and Beard, Gemma Faye and Berryessa, Colleen M},
	year = {2024},
}

@inproceedings{umbachPrevalenceImpactsImageBased2025,
	title = {Prevalence and {Impacts} of {Image}-{Based} {Sexual} {Abuse} {Victimization}: {A} {Multinational} {Study}},
	booktitle = chi,
	author = {Umbach, Rebecca and Henry, Nicola and Beard, Gemma},
	year = {2025},
}

@article{timmermanStudyingOnlineDeepfake2023,
	title = {Studying the online deepfake community},
	volume = {2},
	number = {1},
	journal = {Journal of Online Trust and Safety},
	author = {Timmerman, Brian and Mehta, Pulak and Deb, Progga and Gallagher, Kevin and Dolan-Gavitt, Brendan and Garg, Siddharth and Greenstadt, Rachel},
	year = {2023},
}

@book{sexual2012ethical,
	title = {Ethical and safety recommendations for research on perpetration of sexual violence},
	publisher = {Sexual Violence Research Initiative, Medical Research Council},
	author = {Initiative, Sexual Violence Research and Jewkes, Rachel and Dartnall, Elizabeth and Sikweyiya, Yandisa and {others}},
	year = {2012},
}

@article{flynnSexualizedDeepfakeAbuse2025,
	title = {Sexualized deepfake abuse: {Perpetrator} and victim perspectives on the motivations and forms of non-consensually created and shared sexualized deepfake imagery},
	journal = {Journal of Interpersonal Violence},
	publisher = sage,
	author = {Flynn, Asher and Powell, Anastasia and Eaton, Asia A and Scott, Adrian J},
	year = {2025},
}

@misc{centerfordemocracytechnologyTAKEITAct2025,
	title = {{TAKE IT DOWN} Act {Sign-On} {Letter}: {Concerns} {Regarding} {S}. 146},
	author = {{Center for Democracy \& Technology}},
	month = feb,
	year = {2025},
	url = {https://cdt.org/wp-content/uploads/2025/02/TAKE-IT-DOWN-Sign-On-Letter_21225.pdf},
}

@article{fullerPornWarsSerious2019,
	title = {Porn wars: {Serious} value, social harm, and the burdens of modern obscenity doctrine},
	journal = {Journal of Gender, Social Policy \& the Law},
	publisher = {HeinOnline},
	author = {Fuller, P Brooks and Wagner, Kyla P Garrett and Mazandarani, Farnosh},
	year = {2019},
}

@misc{u.s.supremecourtMarvinMillerState1973,
	title = {Marvin {Miller} v. {State} of {California}},
	author = {{U.S. Supreme Court}},
	month = jun,
	year = {1973},
}

@misc{nationalcenterformissing&exploitedchildrenNationalCenterMissing2026,
	title = {National {Center} for {Missing} \& {Exploited} {Children}},
	url = {https://www.missingkids.org/home?lang=en-US},
	author = {{National Center for Missing \& Exploited Children}},
	year = {2026},
}

@inproceedings{princessacintaqiaStopNonconsensualUse2025,
	title = {Stop the {Nonconsensual} {Use} of {Nude} {Images} in {Research}},
	booktitle = {Proc. of NeurIPS},
    author={Cintaqia, Princessa and Arya, Arshia and Redmiles, Elissa and Kumar, Deepak and McDonald, Allison and Qin, Lucy},
	year = {2026},
}

@misc{pieterhaeckEUSetBan2026,
	title = {{EU} set to ban {AI} nudification apps in wake of {Grok} scandal},
	url = {https://www.politico.eu/article/eu-grok-x-elon-musk-ai-nudification-ban-in-wake-of-scandal/},
	howpublished = {Politico},
	author = {{Pieter Haeck}},
	month = mar,
	year = {2026},
}

@misc{parliamentoftheunitedkingdomDataUseAccess2025,
	title = {Data ({Use} and {Access}) {Act} 2025, {Section} 138},
	url = {https://www.legislation.gov.uk/ukpga/2025/18/section/138},
	author = {{Parliament of the United Kingdom}},
	month = jun,
	year = {2025},
	note = {Issue: c. 18},
}

@misc{huiWhatKnowUK2026,
	title = {What to know about {UK} legal changes aiming to regulate {AI}-generated nude images},
	url = {{https://apnews.com/article/uk-grok-regulation-laws-63406274d1c3040bf4da60460577f34c}},
	journal = {Associated Press},
	author = {Hui, Sylvia},
	month = jan,
	year = {2026},
}

@article{doringExperiencesAIGeneratedPornography2025,
	title = {Experiences with {AI}-{Generated} {Pornography}: {A} {Quantitative} {Content} {Analysis} of {Reddit} {Posts}},
	journal = {Archives of Sexual Behavior},
    author={D{\"o}ring, Nicola and Le, Thuy Dung and Miller, Dan J},
	month = sep,
	year = {2025},
}

@article{lapointePresentFutureAdult2025,
  title={The present and future of adult entertainment: A content analysis of AI-generated pornography websites},
  author={Lapointe, Valerie A and Dub{\'e}, Simon and Rukhlyadyev, Sophia and Kessai, Tinhinane and Lafortune, David},
  journal={Archives of Sexual Behavior},
  pages={1--19},
  year={2025},
  publisher={Springer}
}

@inproceedings{dawoudUndergroundMainstreamMarketplaces2026,
	title = {From {Underground} to {Mainstream} {Marketplaces}: {Measuring} {AI}-{Enabled} {NSFW} {Deepfakes} on {Fiverr}},
	booktitle = {Proc. of {USEC}},
	author = {Dawoud, Mohamed Moustafa and Cuevas, Alejandro and Raman, Ram Sundara},
	year = {2026},
}

@misc{burgessMillionsPeopleAre2024,
	title = {Millions of {People} {Are} {Using} {Abusive} {AI} ‘{Nudify}’ {Bots} on {Telegram}},
	url = {https://www.wired.com/story/ai-deepfake-nudify-bots-telegram/},
	journal = {WIRED},
	author = {Burgess, Matt},
	month = oct,
	year = {2024},
}

@misc{katecongerElonMusksGrok2026,
	title = {Elon {Musk}'s {Grok} {A}.{I}. {Chatbot} {Made} {Millions} of {Sexualized} {Images}, {New} {Estimates} {Show}},
	url = {https://www.nytimes.com/2026/01/22/technology/grok-x-ai-elon-musk-deepfakes.html},
	journal = {The New York Times},
	author = {Conger, Kate and Freedman,Dylan and Thompson, Stuart A.},
	month = jan,
	year = {2026},
}

@misc{organizationfortransformativeworksArchiveOurOwn,
	title = {Archive of Our {Own}},
	url = {https://archiveofourown.org/},
	author = {{Organization for Transformative Works}},
    year = {2026},

}

@inproceedings{riccio2024exposed,
  title={Exposed or erased: algorithmic censorship of nudity in art},
  author={Riccio, Piera and Hofmann, Thomas and Oliver, Nuria},
  booktitle=chi,
  year={2024}
}

@article{cretu2025evaluating,
  title={Evaluating Concept Filtering Defenses against Child Sexual Abuse Material Generation by {Text-to-Image} Models},
  author={Cretu, Ana-Maria and Kireev, Klim and Abdalla, Amro and Obinna, Wisdom and Meier, Raphael and Bargal, Sarah Adel and Redmiles, Elissa M. and Troncoso, Carmela},
  journal={arXiv preprint arXiv:2512.05707},
  year={2025}
}

@phdthesis{austin2023identity,
  title={Identity construction in the furry fandom},
  author={Austin, Jessica R},
  year={2023},
  school={Anglia Ruskin Research Online (ARRO)}
}

@article{peterson2024can,
  title={Can a {Sex-Positive} Sex Education Intervention Reduce {Rape-Supportive} Attitudes? Formative Research, Pilot Randomized Controlled Trial, and Evaluation of {Self-Selection} Bias},
  author={Peterson, Zo{\"e} D and Richmond, Kaylee P and Holt, Laura and Beale, Karen S and Cook, Victoria},
  journal={American Journal of Sexuality Education},
  year={2024},
  publisher={Taylor \& Francis}
}

@article{carmody2005ethical,
  title={Ethical erotics: Reconceptualizing anti-rape education},
  author={Carmody, Moira},
  journal={Sexualities},
  year={2005},
  publisher={Sage Publications Sage CA: Thousand Oaks, CA}
}

@article{sower2023kink,
  title={Kink and {BDSM} Awareness in Sex Offense Treatment},
  author={Sower, Emma and Neal, Bronwyn and Schmader, Alexis},
  journal={Journal of Positive Sexuality},
  year={2023}
}

@article{williams2015moving,
  title={Moving full-speed ahead in the wrong direction? A critical examination of {US} {sex-offender} policy from a positive sexuality model},
  author={Williams, DJ and Thomas, Jeremy N and Prior, Emily E},
  journal={Critical Criminology},
  year={2015},
  publisher={Springer}
}

@article{harling2019influence,
  title={The influence of interviewers on survey responses among female sex workers in Zambia},
  author={Harling, Guy and Chanda, Michael M and Ortblad, Katrina F and Mwale, Magdalene and Chongo, Steven and Kanchele, Catherine and Kamungoma, Nyambe and Barresi, Leah G and B{\"a}rnighausen, Till and Oldenburg, Catherine E},
  journal={BMC medical research methodology},
  year={2019},
  publisher={Springer}
}

@inproceedings{wei2025utterlyillprepared,
author = {Wei, Miranda and Yeung, Christina and Roesner, Franziska and Kohno, Tadayoshi},
title = {"We're utterly ill-prepared to deal with something like this": Teachers' Perspectives on Student Generation of Synthetic Nonconsensual Explicit Imagery},
year = {2025},
booktitle = chi,
}

@article{hearn2007background,
  title={Background paper on guidelines for researchers on doing research with perpetrators of sexual violence},
  author={Hearn, Jeff and Andersson, Kjerstin and Cowburn, Malcolm},
  year={2007},
  publisher={Sexual Violence Research Initiative},
  journal={Sexual Violence Research Initiative}
}

@article{sikweyiya2011perceptions,
  title={Perceptions about safety and risks in gender-based violence research: implications for the ethics review process},
  author={Sikweyiya, Yandisa and Jewkes, Rachel},
  journal={Culture, health \& sexuality},
  year={2011},
  publisher={Taylor \& Francis}
}

@article{braun2006using,
  title={Using thematic analysis in psychology},
  author={Braun, Virginia and Clarke, Victoria},
  journal={Qualitative research in psychology},
  year={2006},
  publisher={Taylor \& Francis}
}

\clearpage
\appendix

\section{Demographics}
% \label{appendix:methods}

% \subsection{Demographics}
\label{appendix:demographics}

\header{Geographic Location}
At the time of their interview, the majority of our participants resided in North America: the United States ($n$=$9$), Canada ($n$=$3$), and Mexico ($n$=$2$). They also resided in the following countries: Australia, Brazil, Chile, China, Colombia, France, India, Indonesia, The Netherlands, the United Kingdom ($n$=$1$ each).

\header{Age}
Two participants were in the 18-21 age group, four participants were in the 22-24 age group, 
twelve participants were in the 25-34 year old age group, 
five participants were in the 35-44 age group, and 1 individual was in the 45-54 age group.

\header{Optional Responses}
Participants were provided open-text response fields for questions on race \& ethnicity, gender, and their sexual orientation.

Slightly over half ($n$=$13$) of our participants self-described as White or Caucasian, including those who self-described as ``White \& Hispanic'', ``Mexican caucasian'', and ``presenting as white''.
Six responded that they are Hispanic and/or Latin American and five responded that they are Asian or South Asian.

Table~\ref{table:demographics} displays participants' responses to optional questions on gender and sexual orientation. In order to present summary statistics, we categorized responses on gender and sexual orientation based on their closest categorical approximation. However, these results do not fully capture the richness of responses provided in self-descriptions.

Lastly, participants could optionally respond if they had a disability (including physical, mental health, neurodivergence). While three preferred not to respond, eleven participants responded ``no'' and ten responded ``yes''.

\begin{table}[!ht]
    \centering
  \begin{tabular}{l|c}
    \toprule
    How would you describe \\ your sexual orientation? & \# Participants \\
    \midrule
     Heterosexual & 16  \\
    Bisexual/Gay/Lesbian/Pansexual/Queer & 7 \\
    No Response & 1 \\
    \bottomrule
    
    \toprule
    How would you describe your gender? & \# \\
      \midrule
    Man & 18 \\
    Non-Binary & 4 \\
    Woman & 3 \\
    No Response & 1 \\
    \bottomrule
  \end{tabular}
    \caption{Gender and sexual orientation. * Several participants identify as multiple genders, and thus,
these counts are not mutually exclusive.}
    \label{table:demographics}
\end{table}

\begin{table}[!h]
\centering
\begin{tabular}{|l|c|}
\hline
\textbf{Category} & \textbf{Count} \\
\hline
Gig work & 8 \\
Science and technology & 7 \\
Other & 6 \\
Arts, culture and entertainment & 5 \\
Communications & 4 \\
Sales & 3 \\
Health and medicine & 3 \\
Architecture and engineering & 2 \\
Education & 2 \\
Business, management and administration & 2 \\
Installation, repair and maintenance & 2 \\
\hline
\end{tabular}
\caption{Occupational background (participants could select more than 1 category).}
\label{table:occupation}
\end{table}

\begin{table}[h!]
\centering
\begin{tabular}{|l|c|}
\hline
\textbf{Education Level} & \textbf{Count} \\
\hline
Did not finish high school & 1 \\
High school graduate (diploma or equivalent) & 4 \\
Some college or Associate degree (e.g., A.A., A.S.) & 7 \\
Trade, technical, or vocational training & 2 \\
Bachelor's degree (e.g., B.A., B.S., B.Eng.) & 7 \\
Postgraduate degree (e.g., Master's, J.D., Ph.D.) & 3 \\
\hline
\end{tabular}
\caption{Educational background.}
\label{table:education}
\end{table}

\end{document}